\documentstyle[11pt,epsfig]{article}

\begin{document}



\begin{titlepage}

\rightline{UTEXAS-HEP-98-13}

\begin{center}

\Large
{\bf A Compact Beam Stop for a Rare Kaon Decay Experiment} 
\normalsize

\vspace{0.25cm}

J.~Belz$^{\rm a,1}$, M.~Diwan$^{\rm b,2}$, M.~Eckhause$^{\rm c}$,
C.M.~Guss$^{\rm a,3}$, A.D.~Hancock$^{c}$, A.P.~Heinson$^{\rm d,4}$,
V.L.~Highland$^{\rm a,5}$, G.W.~Hoffmann$^{\rm e}$, G.M.~Irwin$^{\rm b}$,
J.R.~Kane$^{\rm c}$, S.H.~Kettell$^{\rm a,2}$, Y.~Kuang$^{\rm c,6}$,
K.~Lang$^{\rm e}$, J.~McDonough$^{\rm e,7}$, W.K.~McFarlane$^{\rm a,8}$,
W.R.~Molzon$^{\rm d}$, 
P.J.~Riley$^{\rm e}$,
J.L.~Ritchie$^{\rm e}$, A.J.~Schwartz$^{\rm b,9}$, 
B.~Ware$^{\rm e,10}$, R.E.~Welsh$^{\rm c}$, R.G.~Winter$^{\rm c,5}$,
M.~Witkowski$^{\rm c,11}$, S.G.~Wojcicki$^{\rm b}$
S.D.~Worm$^{\rm *,e,12}$,~and~A.~Yamashita$^{\rm e,13}$

\vspace{0.25cm}

$^{\rm a}${\it Temple University, Philadelphia, Pennsylvania 19122} \\
$^{\rm b}${\it Stanford University, Stanford, California 94309}     \\
$^{\rm c}${\it College of William and Mary, Williamsburg, Virginia 23187}   \\
$^{\rm d}${\it University of California, Irvine, California 92717} \\
$^{\rm e}${\it University of Texas, Austin, Texas 78712}            \\

\end{center}

\begin{center}
{\bf Abstract}
\end{center}       

We describe the development and testing of a novel beam stop for use in
a rare kaon decay experiment at the Brookhaven 
Alternating Gradient Synchrotron.  
The beam stop
is located inside a dipole spectrometer magnet 
in close proximity to
straw drift  chambers
and intercepts a high-intensity neutral hadron beam. 
The design process, involving both Monte Carlo
simulations and beam tests of
alternative beam-stop shielding arrangements, had  the goal of
minimizing the  leakage of particles 
from the beam stop and the resulting hit rates in detectors,
while preserving maximum acceptance for events of interest.  
The beam tests consisted of measurements of rates in drift chambers,
scintillation counter hodoscopes, a gas threshold Cherenkov counter, 
and
a lead glass array. 
Measurements were also made with a set of specialized
detectors which were sensitive to low-energy neutrons, photons, and charged
particles.  Comparisons are made between these measurements and a detailed
Monte Carlo simulation.

\vspace{0.25cm}

\small

\noindent{$^*$Corresponding author.  
Email: worm@fnal.gov.
Phone: (630)840-5011.} \\
\noindent $^1$Present address: Rutgers University, Piscataway, NJ 08855. \\
\noindent $^2$Present address: Brookhaven National Laboratory,
Upton, NY 11973. \\
\noindent $^3$Present address: Cornell University, Ithaca, NY 14853.   \\
\noindent $^4$Present address: University of California,
Riverside, CA 92521. \\
\noindent $^5$Deceased.  \\
\noindent $^6$Present address: Lucent Technologies, Naperville, IL 60566. \\
\noindent $^7$Present address: University of Pennsylvania
Medical Center, Philadelphia, PA 19104. \\
\noindent $^8$Present address: Norfolk State University, Norfolk, VA 23504.  \\
\noindent $^{9}$Present address: University of Cincinnati, Cincinnati,
OH 45221-0011.  \\
\noindent $^{10}$Present address: Caltech, Pasadena, CA 91125.   \\
\noindent $^{11}$Present address: Rensselaer Polytechnic Institute, Troy,
NY 12180. \\
\noindent $^{12}$Present address: Univerisity of New Mexico, Albuquerque,
NM 87131. \\
\noindent $^{13}$Present address: Spring-8, Hyogo 678-12, Japan.    \\
\normalsize

\noindent{Submitted to Nuclear Instruments and Methods.}

\end{titlepage}


\section{Introduction}

Hadronic interactions create difficult shielding problems
in a variety of experimental environments.  Absorbers,
ranging from calorimeters to
passive beam dumps, may be located close to detectors
whose performance can be degraded by leakage
of charged particles, neutrons, or
photons.   
Obtaining reliable {\it a priori} estimates
of the leakage can be important
in the design of an experiment.  Minimizing leakage
under circumstances where the size, shape, and position
of the absorber are imposed by other considerations may
be a goal in the design of the absorber.

In the rare $K^0_L$ decay Experiment~871~\cite{proposal} at 
the Alternating Gradient Synchrotron
(AGS) of the Brookhaven National Laboratory (BNL),  we have addressed
the problem of designing a beam stop for a high intensity
multi-GeV neutral hadron beam.  The location of the beam stop inside
the E871 spectrometer within a few centimeters of tracking chambers,
combined with the fact that the transverse
dimensions of the beam stop were tightly
constrained by the requirement of maintaining good geometrical
acceptance for $K_L^0$ decays,
made this a challenging task.

The beam stop design proceeded in three basic steps:
(1) a preliminary design was arrived at based on general features
of hadronic cascades and on rudimentary simulations of
different shielding options~\cite{SGWmemo};  
(2) extensive beam tests were performed to evaluate
various trade-offs in the design;
and (3) detailed simulations 
were performed and compared to the
beam test measurements~\cite{SWthesis}, 
and were used to further optimize the
beam stop design.
The resulting
design relies on a dense core of tungsten alloy (``heavimet")
surrounded by multiple layers of different materials
to degrade, thermalize, and capture neutrons, and then
to absorb the resulting photons.  
A re-entrant tunnel reduces the backwards-going leakage.
BNL E871 has
subsequently completed  successful  physics runs
in 1995 and 1996.

This paper describes 
the design process, beam tests, and Monte Carlo simulations
that produced a satisfactory beam stop design.
Section~2  describes BNL E871 in more detail and explains the
motivation for placing a beam stop inside the spectrometer.
Section~3 discusses the general considerations leading to
the basic features of the beam stop design.
Section~4 describes the beam tests performed to study
the leakage from the beam stop and its effect on nearby
particle detectors (including drift chambers, scintillation
counters,  a gas threshold Cherenkov counter, and a lead glass
calorimeter).
Section~5 describes the Monte Carlo simulations
performed to study the beam stop and Section~6 compares
the Monte Carlo results to data.
Section~7 summarizes the lessons from this work.

\section{BNL E871}

The goal of
BNL E871 is to
search for the dilepton decay modes of the long-lived neutral kaon $K_L^0$,
such as $K_L^0 \rightarrow \mu^\pm e^\mp$,
$K_L^0 \rightarrow \mu^+ \mu^-$, and $K_L^0 \rightarrow e^+ e^-$,
with
single event sensitivity near 10$^{-12}$. 
Our earlier experiment, BNL E791~\cite{mumu}, 
achieved a single event sensitivity
close to $10^{-11}$ for
these modes,  observed over 700 $K_L^0 \rightarrow \mu^+ \mu^-$ events, 
and set upper bounds on the other two modes.
In order to design an experiment with an order of magnitude improvement
in sensitivity, we had to sample a much larger number of decays and at the
same time improve background rejection.

A larger sample of $K_L^0$ decays was made possible by using a more intense
primary proton beam (which became available with the construction
of the AGS Booster), by increasing the length of the decay volume, and
by increasing the aperture of the spectrometer.
However, operating under these more challenging conditions required a more
efficient trigger to relieve the load on the data acquisition system.
A classical way to make a selective trigger for two-body 
$K_L^0$ decays exploits the two-body kinematics by
arranging the magnetic field(s) in the spectrometer so
the two charged daughters exit the field region with trajectories
approximately parallel to the direction of the parent.
We adopted this general concept owing to its simplicity.

Excellent background rejection requires excellent resolution for the
reconstructed invariant mass of the observed dilepton pair, 
which in turn requires relatively strong magnetic fields to achieve the
required momentum resolution.  In addition, rejecting the background from
$K_L^0 \rightarrow \pi^\pm e^\mp \nu$ where the pion decays 
to
a muon
inside a spectrometer magnet  is greatly improved by using 
two spectrometer magnets in
sequence (to allow comparison of two independently measured
momenta).
Since a transverse momentum ($p_T$) kick of only 220~MeV/c is required to
produce the desired parallelism, two magnets of the same polarity 
in sequence would each need $p_T$-kicks of
110~MeV to achieve this condition, but would operate with
rather low fields and fail
to satisfy our 
resolution requirement.  Therefore, a configuration with two magnets
of opposite polarity is more desirable.  In view of the magnets
available at the time we were planning E871, 
a configuration with a $+440$~MeV/c $p_T$-kick followed by a $-220$~MeV/c
kick seemed attractive.  
This configuration, for those events which exit parallel,
deflects the charged daughters inward (toward the beam)
in the first magnet and outward (to parallel) in the second
magnet, making a so-called ``in-bend" event.
The alternative configuration with 
the smaller $p_T$-kick first would result in
parallel events which were ``out-bends".  This was less
attractive because
the size of the detector downstream of the magnets would
have had to be larger and  the mass resolution would have
been poorer (accepted events would have had smaller opening
angles between the decay products than in the in-bend case, 
resulting in  larger relative uncertainty in the opening
angle).

A potential drawback of this preferred configuration is that 
downstream of the magnets a significant
fraction of the daughter particles  wind up close to the
beam volume or even inside it, making their detection problematical.
A salient feature of these events is that the particles
pass through the first magnet well outside the beam, leaving a
zone through which no particles of interest pass.
Because of this feature,  a compact beam stop
placed in this zone can, in principle, clean up the downstream
environment with only small loss of acceptance.  Consequently,
a constraint on the width of the beam stop (about 40~cm) 
is imposed by the
need to maintain good acceptance.
By eliminating the neutral beam, it not only becomes possible
to detect the particles in that region, but additional
improvement in background rejection can result from the more
reliable
particle identification which is possible in the lower
rate environment created downstream of the beam stop.
The reduced rates come  about because
there
are no longer $K_L^0$ decays along the entire length of the experiment,
nor the inevitable scraping of tails of the neutral beam on detector
frames.

Thus we were led to consider a compact beam stop in the neutral
beam inside the first spectrometer magnet in E871.
However, it was clear that a penalty would result in the form of
higher rates in nearby detectors, owing to leakage from the
beam stop.  {\it A priori} it was not clear whether the penalty
would be prohibitive.
A first attempt to address this was made at the end of the
1990 run of E791 in a short beam test using a beam stop composed
of lead bricks surrounded by polyethylene.
This naive design generated unacceptably high rates, but did not
discourage us from an effort to build an optimal beam stop.
The effort
spanned two years and involved both beam tests and
Monte Carlo simulations.

E871 resides in the B5 neutral beam at the AGS.
The neutral beam is produced by  24~GeV/c protons
in a 12~cm long platinum  target with $3.2 \times 2.5$~mm$^2$ 
transverse dimensions. A series of magnets, lead foils, and collimators
convert photons and sweep away the resulting 
charged particles, leaving a neutral beam
with a solid angle of $4 \times 16$~mrad$^2$ 
(the larger dimension
is the vertical). The centerline of the neutral beam
is oriented at 3.75~degrees 
relative to the proton beam.
The design intensity for E871 is $1.5 \times 10^{13}$ protons on target
over a one second spill, which provides
a $K_L^0$ flux of about $3 \times 10^8$ per spill.
Hereafter we will adopt the unit Tp for
incident proton flux, where
1~Tp = $10^{12}$ protons on target, so that our design intensity was 15~Tp.

The E871 spectrometer, shown in Fig.~1,
consists of a tracking section followed by
scintillator hodoscopes  and  particle identification
detectors.  The tracking section consists of two 
consecutive dipole magnets and
six planes of tracking chambers.  The beam stop is located in the more
upstream of the two dipole magnets.
The four tracking chamber planes nearest the beam stop consist of
5~mm diameter straw drift chambers~\cite{Vienna}
 (two pairs of straw chambers precede 
the beam stop and two pairs
follow it), operated with a fast CF$_4$-based
drift gas~\cite{AnnsNIM}.  Two planes of conventional drift
chambers with 1~cm pitch follow the second dipole magnet.
The magnets are run with opposite polarities tuned to provide a
net $p_T$ kick of about  220~MeV, 
as described earlier.
A pair of scintillation counter hodoscopes, separated by 2.8~m,
follow the tracking section and are the basis for a first level hardware
trigger which requires the two-track parallelism described earlier.  
Particle identification is provided by an atmospheric hydrogen
threshold Cherenkov  counter located between the
trigger hodoscopes, a lead glass array following the final
trigger hodoscope, and finally an instrumented absorber to track muons
and measure their range. 
Data from detectors are digitized in custom high speed front-end 
electronics modules before being fed through a massively parallel readout
system 
into a farm of eight RISC processors, where events were selected in
software for permanent storage.

\begin{figure}
\begin{center}
\mbox{\epsfig{file=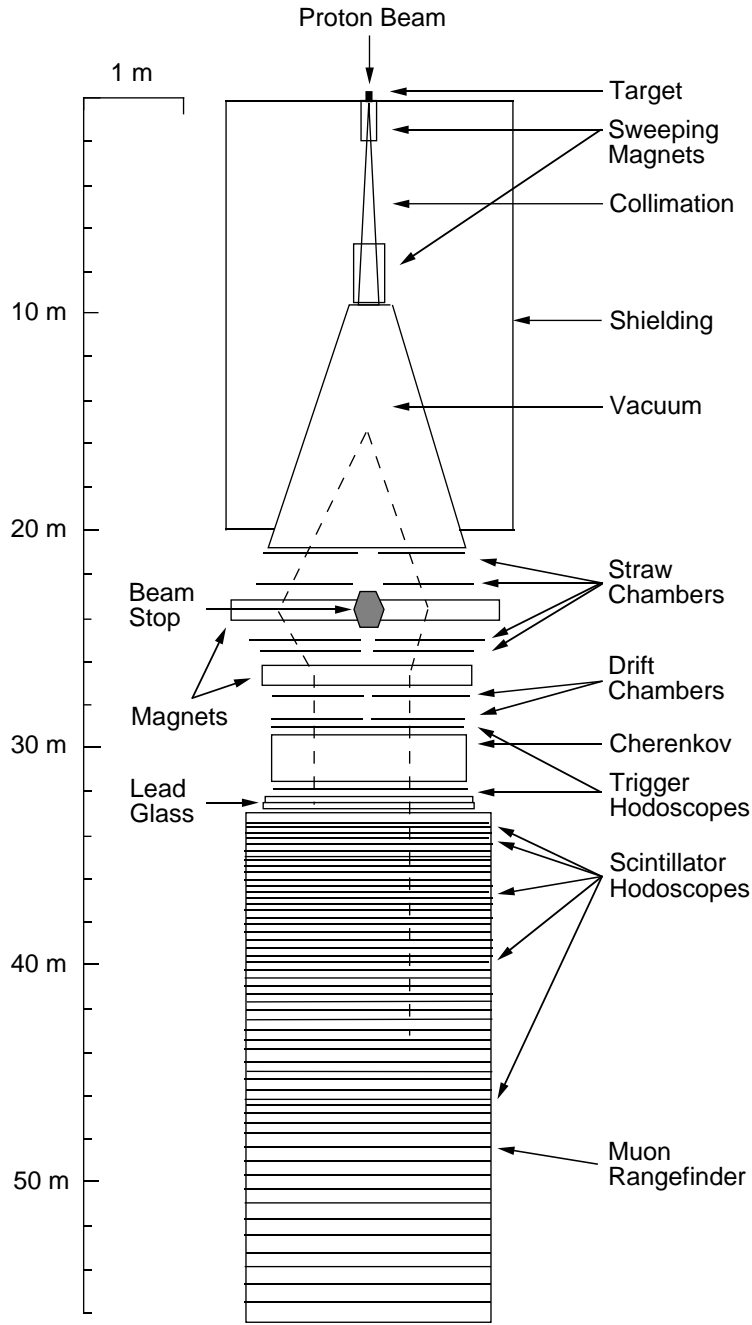,height=500pt}}
\end{center}
\caption{Plan view of the E871 detector.
The dashed lines illustrate trajectories of particles from
a hypothetical $K_L^0 \rightarrow \mu^\pm e^\mp$ decay.}
\label{fig1}
\end{figure}

We are aware of one other fixed-target detector,
that of Fermilab experiment 605~\cite{E605},
which utilized a beam stop
in a somewhat similar arrangement.  
E605 searched for high-mass dimuon pairs and high $p_T$ hadrons
in proton-nucleus collisions.  In that experiment a beam dump inside a
large dipole magnet was well separated from active detectors.
More recently, Fermilab experiments 772 and 866 have carried forward
substantially the same arrangement.

\section{General Considerations for the Beam Stop Design}

A preliminary design for the beam stop was
based on  general considerations, supplemented by
Monte Carlo simulations.
Placing the beam stop inside the first spectrometer magnet
limits its length to 260~cm.   Because it is located inside a magnet,
it should be constructed of
nonmagnetic materials in order 
to avoid distorting the
magnetic field. 
Also, the forces on magnetic materials would have required a more
complicated mechanical design.
The width of the beam stop is limited by the
need to maximize acceptance for two-body $K_L^0$ decays. 

The beam stop needs a central high-density core 
with a short
hadronic interaction length in order
to concentrate the major fraction of the hadronic cascade in a 
small volume.  This assures that most of the particles produced
in the development of the hadronic cascade will
either interact or be ranged out before reaching the 
periphery of the beam stop.  In addition, such material
helps to minimize the number of decays of secondary pions
to muons.
The
inelastic collisions that are required to stop the incident neutrons
produce a large number of low energy protons, neutrons, and photons as
the target nuclei release the energy absorbed in the collisions. 
Protons and photons have  short ranges in the dense hadronic
absorber, but the  low energy neutrons, with
energies around 1~MeV, can have long interaction lengths and tend
to lose  little energy in each interaction. 
 
Candidate materials 
for the dense core were tungsten
and depleted uranium because of their short nuclear interaction
lengths.
Tungsten has a nuclear interaction length of 9.6~cm
and  the added benefit of a large inelastic neutron
cross section down to 1~MeV. Depleted uranium has a slightly longer
nuclear interaction length, 10.5~cm, but is less expensive. It also has
a large inelastic neutron cross section, although much of this
is due to fission processes which produce additional
low energy neutrons.  
A study of neutron leakage using the {\small HETC} hadronic 
cascade Monte Carlo
and {\small MORSE} neutron transport code,
which are part of the {\small CALOR89}~\cite{calor} package,
indicated that approximately a factor of three
more low-energy neutrons would  escape from 
a beam stop consisting of a uranium core surrounded by two inches
of polyethylene than from a beam stop consisting of a 
tungsten core equivalently shielded.
Tungsten was chosen for the core material.   Because tungsten
is expensive, it
was supplemented with copper in some regions
to provide additional attentuation
of the hadronic shower at a lower cost.

It is necessary to thermalize and capture the neutrons which escape
from the core material.
  The most effective method to thermalize low energy neutrons is
elastic scattering on free protons (for example, 
in a hydrogen-dense material
such as polyethylene). 
Once thermalized, the neutrons
are usually captured by a nucleus, a process that often results in the
production of a photon with  energy in the MeV range. 
For example, neutron capture
on hydrogen  
produces a 2.2~MeV photon through the process
$^1$H($n,\gamma$)$^2$H (i.e., $ n +  p \rightarrow d + \gamma $).
Photons with this energy are  difficult to stop;
the attenuation length in lead is approximately 2~cm. By doping the
hydrogen-dense thermalizing material with boron or lithium, 
one can cause the neutron to be
captured in a reaction that produces a lower energy photon, 
such as
$^{10}$B($n,\alpha$)$^7$Li$^*$, or  no photon, 
such as $^6$Li($n,\alpha$)$^3$H. 
The boron reaction $^{10}$B($n,\alpha$)$^7$Li$^*$
emits a 0.477 MeV photon, which has an
attenuation length of about 0.5~cm in lead,
making its absorption more practical. 
In
both cases the range of the $\alpha$-particle is very short so it does not
escape the beam stop. 

  Several materials were considered for the parts of the beam stop that
moderate the secondary neutrons. Pure polyethylene has a very high
hydrogen concentration but generates many 2.2~MeV photons, as discussed
above. Borated polyethylene retains most of the hydrogen density while
generating photons that are easier to attenuate. Lithium-doped
polyethylene generates no photons but has a much lower hydrogen density,
so it is less effective at thermalizing neutrons. 
We  attempted to optimize the design by utilizing different
types of polyethylene in different locations.
For instance, 
in our design the polyethylene along the sides of the heavimet core
was only slightly borated (0.5\%), in order to maintain high hydrogen density.
Thin sheets of a highly borated  material
surrounded this type of polyethylene.
To absorb the photons from neutron capture reactions, a  layer of
lead lined the outside of the
beam stop.  

  The preliminary beam stop design is
shown in Fig.~2.
It consisted of a nonmagnetic heavimet core (94\% tungsten, 10.1~cm
interaction length), 
30.3~cm wide at its widest point and
110~cm long. The heavimet core was 70~cm high and had 7.8~cm of copper
plus 12.7~cm of polyethylene above and below it.  The heavimet consisted
of about 200 machined bricks, in a range of sizes, that were manually
stacked into place.
Downstream of the heavimet
was 41~cm of copper and 20~cm of polyethylene. 
At the upstream end, there was a tunnel that
the neutral beam entered before striking the tungsten core.  Heavimet
formed  a 25~cm long portion of the tunnel.  
The tunnel
was extended further upstream by
a 12~cm wide $\times$ 50~cm high $\times$ 50~cm long section
made of 5\% borated polyethylene with layers of 1.3~cm thick lead and
a section of lithium-doped polyethylene. The
sides of the heavimet core
were covered with 4.5~cm thick 0.5\% borated polyethylene. 
On both sides of the polyethylene were 3~mm sheets of 
{\small FLEX/BORON}~\cite{rxe}, 
a highly borated (25\%) silicone material. 
Finally,  0.64~cm of lead formed the outer layer of the beam stop. 
The compositions of the
materials used in the beam stop, 
which in the final design also 
included borated zirconium hydride polyethylene,
are given in Table~\ref{table:mater}.

\begin{figure}
\begin{center}
\mbox{\epsfig{file=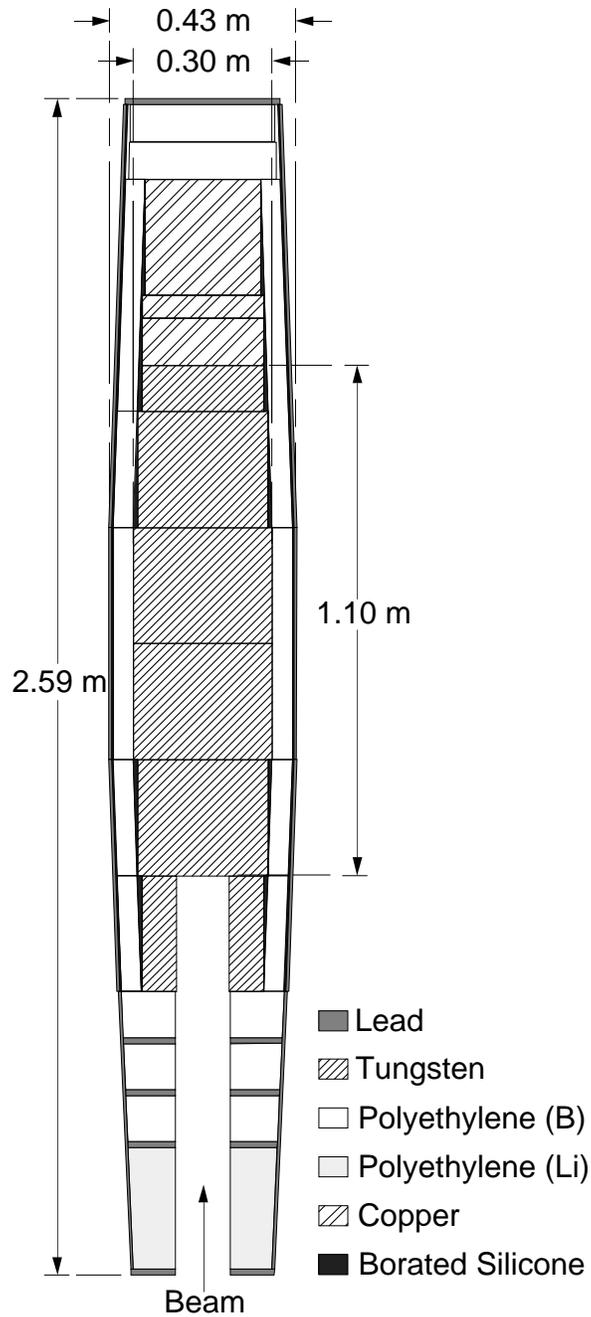,height=500pt}}
\end{center}
\caption{Horizontal cross section view through the center
of the initial beam stop design, described in the text.
A thin layer of {\small FLEX/BORON} surrounded the borated polyethylene.}
\label{fig2}
\end{figure}

\begin{table}
\begin{center}
\begin{tabular}{ c  c l r }
\hline
 Material & Density      & Elemental   &    \\
          & (g/cm$^3$)   & Composition &(\%)\\ \hline \hline
Tungsten Heavimet  &17.95& Tungsten &94.20  \\
                   &     & Nickel   & 4.35  \\
                   &     & Iron     & 0.85  \\
                   &     & Cobalt   & 0.50  \\
                   &     & Copper   & 0.10  \\ \hline
`Stiff' Lead       &11.4 & Lead     &97.    \\
                   &     & Antimony & 3.    \\ \hline
Copper             & 8.96& Copper   &99.9   \\ \hline
Borated Zirconium  & 3.67& Zirconium&85.0   \\
Hydride Polyethylene&    & Carbon   & 8.9   \\
                   &     & Hydrogen & 3.4   \\
                   &     & Oxygen   & 2.2   \\
                   &     & Boron    & 0.5   \\ \hline
{\small FLEX/BORON} & 1.64& Silicon  &26.9   \\
                   &     & Boron    &25.4   \\
                   &     & Oxygen   &24.2   \\
                   &     & Carbon   &20.1   \\
                   &     & Hydrogen & 2.8   \\ \hline
Lithiated Poly     & 1.06& Carbon   &83.7   \\
                   &     & Hydrogen & 8.8   \\
                   &     & Lithium  & 7.5   \\ \hline
Borated Poly, 0.5\%& 0.84& Carbon   &83.1   \\
                   &     & Hydrogen &14.2   \\
                   &     & Oxygen   & 2.2   \\
                   &     & Boron    & 0.5   \\ \hline
Borated Poly, 5.0\%& 0.93& Carbon   &61.2   \\
                   &     & Oxygen   &22.2   \\
                   &     & Hydrogen &11.6   \\
                   &     & Boron    & 5.0   \\ \hline
\end{tabular}
\end{center}
\caption{ \label{table:mater}
Shielding
materials used in the E871 beam stop.  Elemental composition is given
in percent by weight.}
\end{table}

\section{Beam Tests}

A series of 
beam tests were performed during both the 1991 and 1992 AGS running
periods  to study various trade-offs in the beam stop
design and ultimately to assess the viability of this aspect of
E871.  
The tests were conducted in the actual neutral
beamline planned for the experiment (AGS~B5), 
which was the same beamline used by
our previous experiment E791.  At that time the full E791
spectrometer was still intact.  
During the 1991 run, when the most extensive tests were
performed
(including those for which data are reported here), the proton beam
was incident on 
a copper target at a 2.75~degree production angle
rather than the final E871 configuration (a platinum
target at 3.75~degrees).

The increase in targeting angle
reduced the total hadron flux 
(mostly neutrons), while preserving most of the kaon flux.
While this effect is expected, it was also clearly visible
in our data.  We were able to study the angular distribution of
hadrons in the beam 
by studying the position of interactions of beam hadrons
in the vacuum window at the end of the decay volume.
These interactions were identified by 
reconstructing the emerging tracks using our drift chambers.
The $y$-coordinate of the interaction in the window
provided a direct
measurement of the production angle, owing to the fact that
our beam was formed by dumping the primary protons 
with a (downward) vertical deflection and because our collimators
accepted a narrow horizontal slice but a large vertical range
(effectively about 12~mrad centered at 2.75~degrees).
Over this 12~mrad range we observed a drop of 36\% in the 
number of window interactions from the bottom (smaller angle) to the
top (larger angle).
Similarly, we studied the production angle of kaons by 
reconstructing the vertex position of
kaon decays upstream of the vacuum window.  
Kaon flux was almost flat over this angular range.

Earlier E791 
measurements made with the copper target at the 2.75~degree
production angle
indicated there were roughly 
$3.7 \times 10^8$ neutrons in the beam for every
$10^{12}$  incident protons, with  
neutral kaons comprising four to eight percent of the hadrons in the beam.
The neutron flux estimate~\cite{neutr} was based on foil activation studies
and a series of measurements which counted interactions in different
thickness absorbers of a few different absorber materials.

The beam stop studies proceeded in several stages.  In particular,
after the beam stop was partially assembled, 
hit rates in various detectors were measured.  
Then, as  increments in the assembly
were made, the measurements were repeated.
Also, in reaction to data analysis that was going on
in parallel, additional variations to the beam stop configuration were
often tested.
In all, measurements were made for about 95 configurations, although
many of these amounted to small variations on others.
Some preliminary results of these tests have been described
previously~\cite{MDpreprint}.

During the tests,
proton intensities ranged from 0.3 to 2.0~Tp/spill. 
Three types of measurements were made:

\begin{itemize}

\item For all beam stop configurations, we measured the rates in the E791
detectors since those detectors were a reasonably good approximation to the
detectors planned for E871. 
We used a pseudo-random trigger, provided by a 
pulse generator running at a fixed rate (1~kHz), to collect data from
the detector. These triggers, by being 
independent of the time structure of the beam, provided a means of 
effectively randomly sampling
detector activity.
The hit rates in all detectors except the lead glass were
calculated from TDC information.
The hit rates and the radiation dose in the lead glass were
calculated from ADC information.

\item For a reduced set of beam stop configurations,
we made measurements of the neutron, photon, and charged
particle leakage from the beam stop using Bonner spheres,
$^3$He counters, 
liquid scintillators, 
and a high purity germanium
detector.   These detectors 
used a separate data acquisition system 
to read out 
pulse height and timing information for each event. Triggers were
generated from signals provided by the detectors themselves.
A sample of
the events was analyzed in real time to monitor the performance and
stability of these detectors. The events
were written to tape for offline analysis. 

\item For the same
reduced set of beam stop configurations, we measured rates in a
small test drift chamber and four Cherenkov 
photomultiplier tubes (from
the E791 Cherenkov 
counter), that were shielded with lead and polyethylene.
These detectors were read out for the 
same pseudo-random triggers described above
using the E791 data acquisition system.

\end{itemize}

In the following discussion, we will present results from measurements 
made in a
small selected subset of the tested
beam stop configurations.  
The configurations upon which we focus were chosen primarily
because they have
significant differences from each other 
and exhibit the effects of different shielding
configurations upon the leakage.  
A complicating factor in interpreting the
results is that interactions of neutral beam particles
in air upstream of the beam stop  were a significant source of detector
rate.  The interactions in air 
were significantly reduced in some configurations
by the use of helium bags.  As a result, in addition to changes in the
beam stop, it was also necessary to pay attention to whether the helium
bags were present or absent for a particular configuration.  
Because of the requirement of practical access to the beam stop, the
helium bags were absent for most 
configurations for which data was taken.  For completeness, we include below
a configuration which differs from another only with respect to the
use of the helium bags, so that the magnitude of the effect of interactions
in air can be seen.

The configurations discussed in the remainder of this
paper are:

\begin{enumerate}

\item A baseline configuration with no beam stop.  In particular, this
configuration is the normal data-taking configuration of BNL E791 and
included the placement of helium bags in all beam regions downstream
of the vacuum decay region and  between all drift chambers.

\item A purely metallic beam stop.  In this configuration, both the tungsten
heavimet and copper were installed, but no polyethylene or other
neutron degrader or absorber was present.  For this configuration,
helium bags were not in place.

\item The full preliminary beam stop, as described in the previous section
and shown in Fig.~2.
Helium bags were not in place.

\item Extra shielding.  In this configuration, an additional 2.5~cm of
polyethylene was added to the sides of the beam stop 
and an extra 0.64~cm of lead was added to the 
outer layer, as compared to
the configuration
shown in Fig.~2. Helium bags were not in place.

\item Extra shielding with helium bags in place.  For this configuration,
the beam stop was the same as in configuration~4.

\end{enumerate}

Data in each of these configurations was taken with and without a magnetic
field in the region of the beam stop.  That is, the dipole 
magnet in which the
beam stop was located was on for about half of the data in each
configuration and off for the other half.  The magnetic field value
was limited to 0.85~T by the power supply available at the time.
An additional power supply was added later making it possible to
achieve the desired field for E871, which was 1.4~T.

\subsection{\normalsize \it Beam Flux Monitoring}

While it was not necessary to measure the flux in the neutral
beam absolutely, it was important to have a relative normalization
so that data taken at different times and with different beam
stop configurations could be compared.
The results described here used two independent normalization
methods to compare the rates
between different beam stop configurations. The first was based on
counting  $K_L^0$ decays in the vacuum decay
region, which to an excellent approximation should track the
total neutral flux in the beam.
This was accomplished by
reconstructing two tracks emerging from
a common vertex in the data taken with the pseudo-random
trigger. 
Tracks were reconstructed separately in the horizontal ($x$-view)
and vertical ($y$-view) 
planes from the hits in the upstream-most
two drift
chambers.  A vertex, or ``V'', was 
counted if the $z$-position (position along the beam direction)
of the vertices calculated independently using the $x$-view and 
$y$-view tracks coincided  within 7~cm.
In addition, the coordinates of
the intersection point were required to be within the neutral beam
volume. The limitation of this normalization method was
statistics since typically only about 600~V's were found 
in the data
for
each configuration.

  The second normalization method used a triple-coincidence 
scintillator telescope 
that viewed the production target 
at a 90~degree angle. 
The telescope normalization had small statistical uncertainties, but did
not measure the number of secondary particles that were accepted into
the solid angle of the experiment. 
The telescope normalization had a systematic uncertainty of 
about 20\% because
of variations in beam steering onto the target.

  The V normalization could be used only for data taken with the E791
data acquisition system. Therefore, 
the specialized neutron and photon detectors had to rely on 
the telescope
normalization.  Comparisons of the rates between the two detector systems
necessarily used the telescope normalization.

\subsection{\normalsize\it E791 detectors}

While the BNL E791 spectrometer differed in many details from the
planned E871 spectrometer, it was nonetheless
similar in many respects and provided a powerful tool
in studying the effects of the beam stop on particle detectors.
The E791 apparatus is shown in Fig.~3.
It has been described in more detail elsewhere~\cite{mumu}. 
The tracking section
consisted of two consecutive dipole magnets and five conventional
drift chamber planes (two pairs of chambers upstream of the first magnet,
one pair between the magnets, 
and two pairs of chambers downstream of the second
magnet).
Downstream of the final drift
chamber, a finely segmented scintillator hodoscope, a gas threshold 
Cherenkov 
counter, another hodoscope, and a  lead 
glass array followed in sequence.  A meter of steel was then followed
by a scintillator hodoscope for muon identification
and an instrumented range stack for muon range measurement.
The detector was fully operational during the beam tests of the beam
stop. 
However, in order to permit frequent accesses to both sides of the beam stop,
one member of each pair of chambers immediately upstream and downstream of the
beam stop were removed, as indicated in Fig.~3 by the dashed lines.  
In particular,
the right-side
chamber immediately upstream of the
the beam stop and the left-side chamber immediately downstream of
the beam stop were not in place for most of our tests. 

\begin{figure}
\begin{center}
\mbox{\epsfig{file=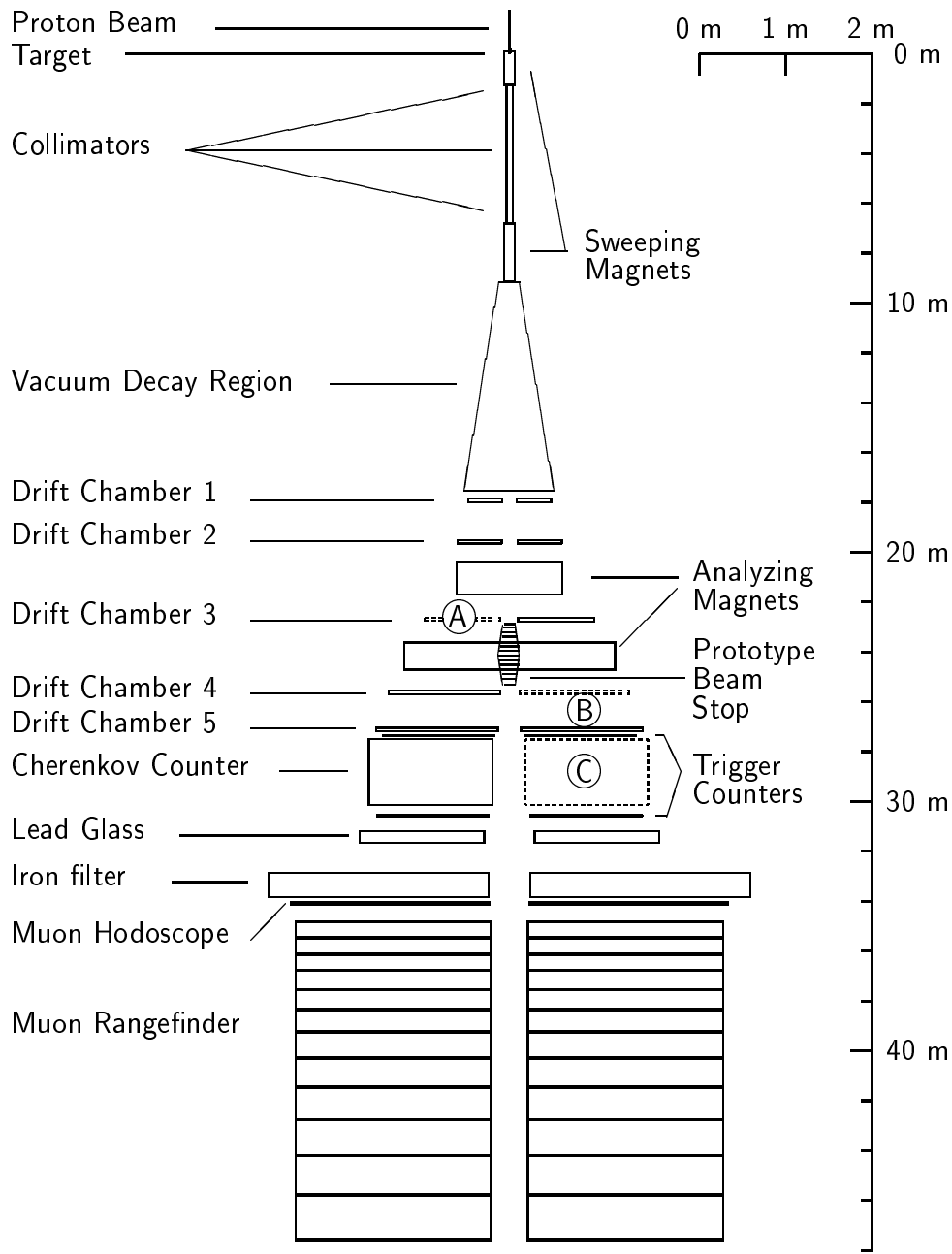,bbllx=130,bblly=160,bburx=502,bbury=690}}
\end{center}
\caption{Plan view of the E791 detector 
showing the position of
the beam stop.
Detector elements removed during these tests are indicated by dashed lines.
The positions indicated by the circled letters $A$ and $B$ are where the
specialized neutron and photon detectors were located.
The circled letter $C$
indicates the location of a test setup for
Cherenkov phototubes.}
\label{fig3}
\end{figure}

For all the tests, the upstream E791 dipole magnet was
off (i.e., zero field), since it was to be removed for
E871.  
The beam stop was located in the second dipole magnet in the E791
setup.  The same magnet in the same location houses the beam stop in
E871.

The hit rates in the E791 detectors for the five previously defined beam stop
configurations are given in Table~\ref{table:rates} 
for data
with the downstream spectrometer magnet 
both at 0~T (magnet off) and at 0.85~T.
Hit rates are given in units of
``hits per V",  the number of detector
hits normalized to the number of kaon-decay vertices observed in the 
same data sample.  This unit takes into account variations in the
incident beam flux and is natural for comparing hit rates
between different data sets and beam stop configurations.
The conversion to hits per unit time per proton on target
can be made by multiplying the
hit rate per V by the factor 25,000 Hz/Tp.  That is, one hit per V 
corresponds to approximately 25~kHz per $10^{12}$ protons on target.
However, since the size and number of channels is not the same for
each detector shown in Table~\ref{table:rates}, only comparisons of
rates in the same detector for different beam stop configurations are
meaningful.

\begin{table}
\begin{center}
\begin{tabular}{ l  c  c  c  c  c  c  c  c  c  c  c }
\hline 
 & \multicolumn{11}{ c }{Beam Stop Configuration} \\ 
Detector &   1  &  2    & 3 & 4 &  5 & \qquad & 1 & 2 & 3 & 4 & 5 \\ \hline
 & \multicolumn{5}{ c }{Magnet Off} & \qquad & 
\multicolumn{5}{ c }{Magnet at 0.85 T} \\ 
DC1  & 120 & 180 & 160 & 140 & 140 & \qquad & 110 & 180 & 160 & 150 & 130 \\ 
DC2  & 160 & 370 & 280 & 250 & 170 & \qquad & 120 & 370 & 280 & 260 & 160 \\
DC3-left  &  220  &  2800  &  1100  
     &  720  &  520  & \qquad & 250 & 2600 & 1000 & 800 & 520 \\ 
 & \multicolumn{11}{ c }{(Beam Stop location)} \\
DC4-right & 280 & 2000 & 900 & 600 & 520 & \qquad & 150 & 1700 & 620 & 
 490 & 420 \\ 
DC5  & 300 & 920 & 430 & 290 & 260 & \qquad & 190 & 850 & 430 & 300 & 260 \\ 
TSC1 & 140 & 820 & 300 & 210 & 190 & & 95 & 850 & 310 & 230 & 180 \\ 
Cherenkov-right & 
                     33 & 58 & 24 & 18 & 16 &
                  &  26 & 56 & 23 & 17 & 13 \\ 
TSC2 & 170 & 290 & 120 & 87 & 77 &  & 140 & 290 & 120 & 89 & 69 \\ 
Lead Glass \qquad & 140 & 83 & 65 & 65 & 49 & & 130 & 72 & 54 & 52 & 37 \\ 
MHO & 110 & 84 & 67 & 74 & 53 & & 110 & 77 & 59 & 62 & 45 \\ \hline
\end{tabular}
\protect 
\caption{ \label{table:rates}
Detector hit rates (hits per V) for the five beam stop
configurations described in the text. The rates are the total number of
hits per side averaged over the left and right sides, except in the cases
of DC3 and DC4, where one side was removed to permit access to the beam
stop, and of the Cherenkov counter, where one side was removed to make
room for a phototube test setup.
The detectors are drift chambers (DC1--DC5), trigger scintillation
counters (TSC1, TSC2), Cherenkov counter, lead
glass array, and muon hodoscope (MHO). One hit per V corresponds to
approximately 25 kHz at a beam intensity of 1~Tp/spill.
Uncertainties of typically 5\% apply to these rates, owing primarily
to the statistical uncertainty on the V normalization.
} 
\protect
\end{center}
\end{table}

A comparison
of hit rates in configuration~2  to
configuration~1 shows that
the rates in the  drift chambers
rose sharply after the beam stop
was installed, especially in the closest chambers (DC3 and DC4).  
The increase for the chamber immediately upstream of the
beam stop (DC3-left) was more than an order of magnitude.  However, in some of the 
configurations with shielding, hit rates in that chamber are reduced to
roughly double the rate with no beam stop.  In other detectors, the
effects of shielding are even more favorable.
The hit
rates in the detectors well downstream of the beam stop are quite low
compared to the rates with no beam stop.  For example, the hit rate in the
downstream trigger hodoscope (TSC2) is about half the rate with no beam
stop. 
The effects observed are similar for both magnet-on and magnet-off data.
Of course, the magnet-on situation corresponds more closely to the
E871 setup.

\subsection{\normalsize\it Neutron and photon detectors}

We used an additional set of detectors to investigate the leakage
of neutrons, photons, and charged particles
from the beam stop. The
neutron detectors were Bonner spheres, $^3$He
counters, 
and liquid scintillators. The liquid scintillators
were also used to measure the photon flux and a high purity germanium
detector was used to identify photons from neutron capture. Charged
particles were identified by requiring a coincidence between a liquid
scintillator and thin plastic scintillators.  
Owing to physical constraints imposed by the E791 spectrometer,
these special detectors could only be located in certain positions.
For the results reported here, the detectors were in one of two positions:
(a)  upstream of the beam stop 
near the position
normally  occupied by DC3-right (i.e., the beam-right
drift chamber immediately upstream of the beam stop;  note that in 
Table~\ref{table:rates} rates are given only for the left-side chamber,
DC3-left, since the right-side chamber was removed for most of these tests);
and (b) downstream of the beam stop approximately halfway
between drift chambers 4 and 5 (DC4 and DC5) on the 
beam-left side.  These positions are indicated in Fig.~3 by the circled
letters $A$ and $B$ for the upstream and downstream positions, respectively.

  Bonner spheres~\cite{refbs} were used to
measure the total neutron flux over a large energy
range. Six polyethylene spheres of different sizes
were placed around a 4 mm
diameter, 4 mm long LiI crystal that detected the recoiling $\alpha$ and
triton
produced when slow neutrons were captured on $^6$Li 
(i.e., $^6{\rm Li} + n \rightarrow \alpha + ^3{\rm H} + 4.8~{\rm MeV}$).
Each sphere (with
diameters of 5.1, 7.6, 12.7, 20.3, 25.4, and 30.5~cm) 
had a known efficiency as a function of neutron energy. Measurements
with the  six
spheres along with  measurements made with (i) the 5.1~cm sphere
covered with cadmium to absorb thermal neutrons, (ii) the LiI detector
covered with cadmium (to absorb thermal neutrons)
but no polyethylene sphere, and (iii) the LiI detector
with no covering at all, provided a total of 
nine points that were used as the input
to the unfolding code {\small BUNKI}~\cite{refbunk}. The output of the
unfolding included the neutron flux (neutrons/cm$^2$) for 30~energy
bins spanning a range from  231~MeV down to thermal neutran energies. 

The pulse height spectra from the LiI
scintillator shown in Fig.~4 indicate that the measurements had very little
background. These spectra represent the test conditions with the
highest and lowest rates in the LiI scintillator. The highest rate was
recorded when the scintillator was covered with a 12.7 cm diameter
polyethylene sphere and the lowest rate occurred when the scintillator
was covered with cadmium but no polyethylene. 
The small difference in the peak positions was caused by a DC-offset
due to the factor of ten difference in rate. 
The long tail below the peak seen in the measurement with the
polyethylene sphere was caused by events in which  neutron capture
occurred so near the edge of the scintillator that the full energies of
the recoiling nuclei were not measured. 
The negative slope in the tail below the peak seen in
the low rate measurement was due to charged particles.
To provide meaningful numbers as
input to the unfolding code, the counts due to minimum ionizing
particles were subtracted from each measurement.

\begin{figure}
\begin{center}
\mbox{\epsfig{file=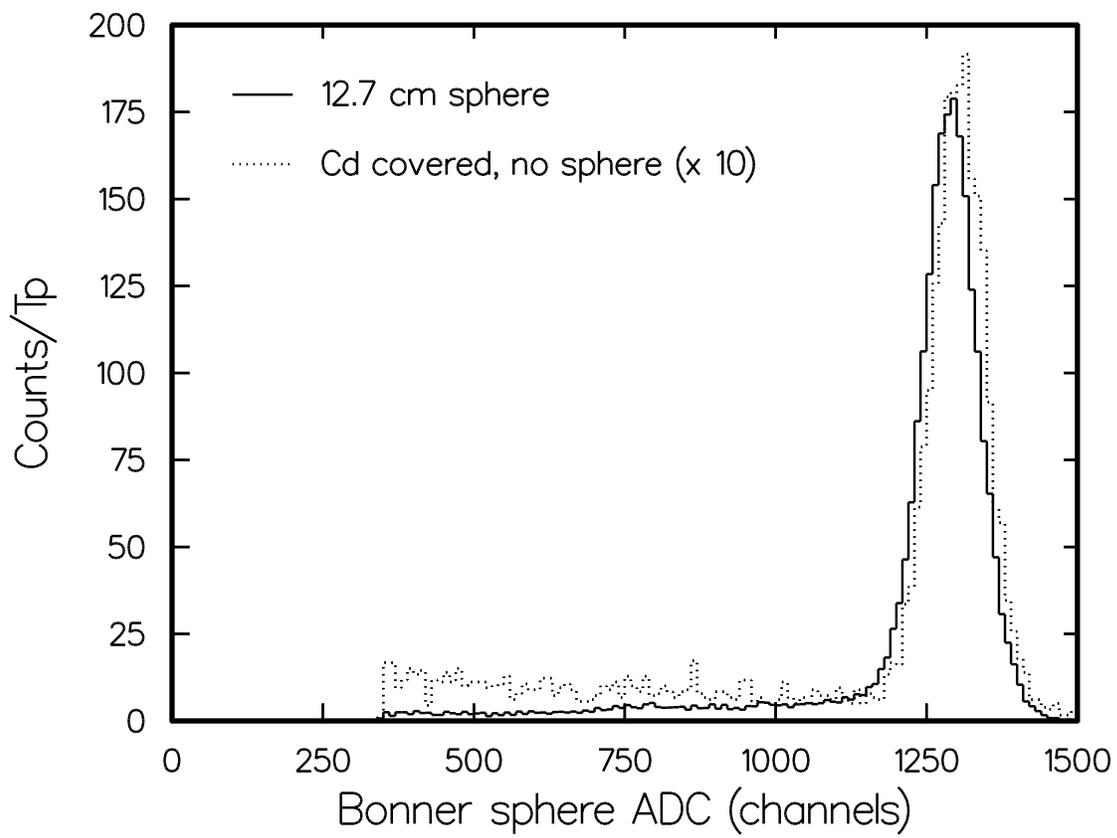,angle=270,width=580pt}}
\end{center}
\caption{ADC spectra from the LiI scintillator used in the
Bonner sphere measurements.}
\label{fig4}
\end{figure}

  The results of the unfolding for four different beam stop
configurations are shown in Fig.~5 for  positions upstream
and downstream of the beam stop. 
One noticeable feature of all the
neutron spectra with the beam stop in place was the low flux 
around 100~eV. 
We interpret this feature as an artifact of the unfolding procedure,
rather than as a physically meaningful.
This feature has previously 
been observed by others~\cite{refbunk} in different situations and is probably 
due to
the fact that no Bonner sphere had a response function 
that peaked in this energy
range, and thus no measurement strongly constrained the unfolding
results. The unfolding procedure had several degrees-of-freedom since it
produced the neutron flux in 30 energy bins with only 9 points of input
data. 

\begin{figure}
\begin{center}
\mbox{\epsfig{file=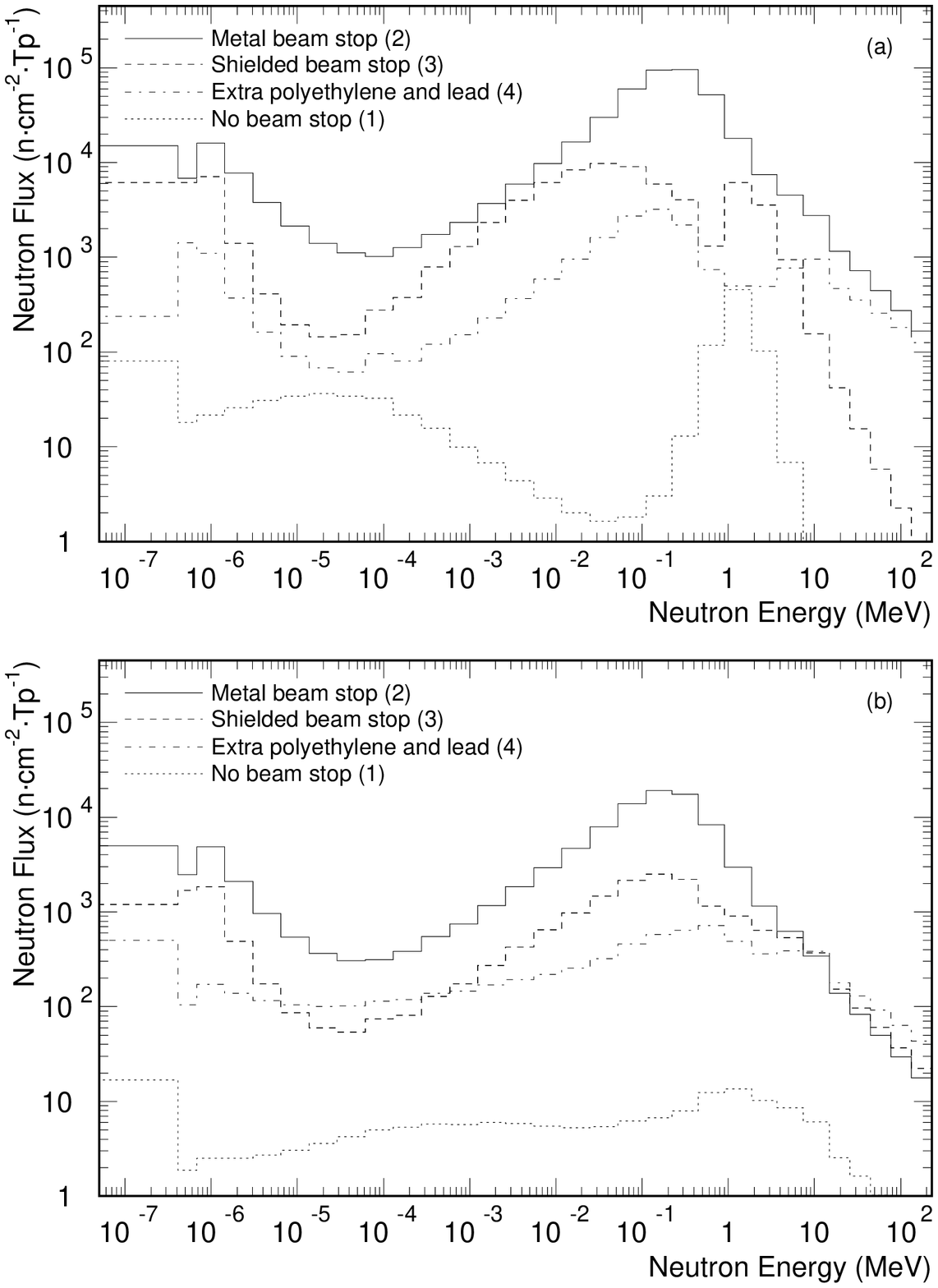,height=500pt}}
\end{center}
\caption{Neutron energy spectra from unfolded
 Bonner sphere measurements
for four beam stop configurations upstream (a) and downstream (b)
of the beam stop.  The numbers in the labels refer to the configurations
defined in the text.}
\label{fig5}
\end{figure}

  The two $^3$He counters~\cite{refhe}, 
which detected only low energy neutrons, provided
a check of the Bonner sphere results. The $^3$He counters had
an active volume
2.5~cm
in diameter and 15~cm long, and were filled to a pressure of 4~atmospheres. 
These gas counters detected slow neutrons via
$^3$He($n,p$)$^3$H and were used in pairs; one was covered with 0.05~cm
cadmium (which stopped neutrons with
energies below about 0.1~eV)
and the other was left unshielded. 
The flux calculated from the Bonner sphere measurements
was used as input to 
calculate the rates expected in the unshielded $^3$He counter
and in the $^3$He counter covered with cadmium. 
The ratios of measured $^3$He counter rates to those
predicted by the Bonner sphere results 
are shown in Table~\ref{table:He3table}.  They exhibit
variations larger than the 20\% uncertainty in the
normalization, but typically agree at the  factor of two level.

\begin{table}
\begin{center}
\begin{tabular}{ l  c  c  c  c }
\hline
Configuration & 1 & 2 & 3 & 4  \\
         & No beam & Tungsten & Shielded  & Extra   \\
         &   stop  &   only   & beam stop & poly, Pb \\ \hline
 & \multicolumn{4}{ c }{upstream position}  \\ 
\ \ \ $^3$He without Cd & 0.6 & 0.9 & 0.9 & 1.9 \\ 
\ \ \ $^3$He with Cd    & 0.8 & 0.6 & 0.8 & 1.1 \\ 
 & \multicolumn{4}{ c }{downstream position}  \\ 
\ \ \ $^3$He without Cd & 2.1 & 0.9 & 0.8 & 1.0 \\ 
\ \ \ $^3$He with Cd    & 2.2 & 0.3 & 0.5 & 0.9 \\ \hline
\end{tabular}
\end{center}
\protect \caption{
The ratio of the  rates in the $^3$He counters to
the rates expected, based on calculations using  the Bonner sphere results
as input. 
}
\label{table:He3table}
\protect
\end{table}

  For neutron energies between 1 and 10~MeV, the neutron flux
measured by the Bonner spheres was checked using the liquid
scintillators.
The 5~cm diameter and 5~cm long liquid scintillator detected
neutrons via ($n,p$) elastic scattering and could identify the neutron-induced 
counts by using a pulse shape discrimination
method~\cite{refpsd}. The scintillation 
produced by neutron interactions had
longer decay times than that from minimum ionizing particles due to
the highly ionizing nature of the 
recoiling protons. 
Pulse shape discrimination from the liquid
scintillators, as shown in Fig.~6 and Fig.~7,  separates neutrons and
photons above an equivalent electron energy of about 350 keV, which
corresponds to ADC
channel 50 in Fig.~6. Between 150 keV and 350 keV the discrimination
between photons and neutrons was less clear, and below 150 keV there was
no separation at all. A 900 keV proton, from ($n,p$) scattering for
example, produces approximately the same amount of scintillation light
as a 150 keV electron and therefore the liquid scintillators identified
only those incident neutrons with energies above 900 keV. 

\begin{figure}
\begin{center}
\mbox{\epsfig{file=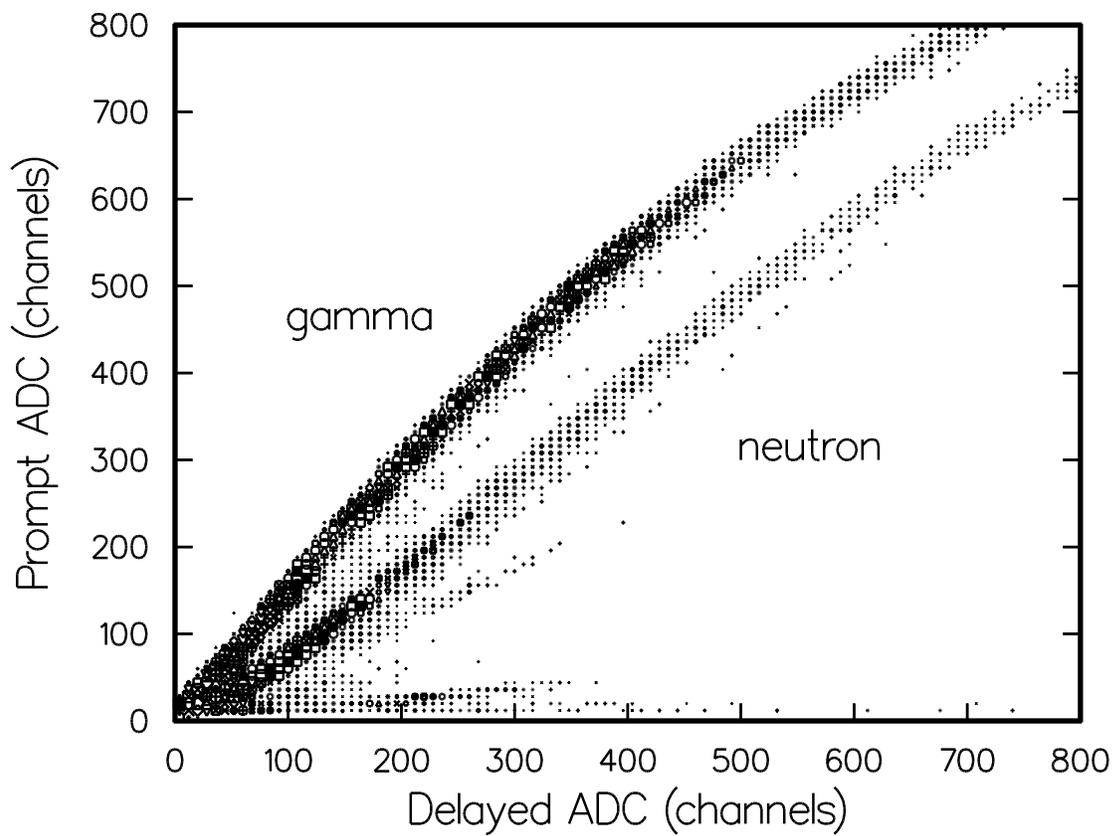,angle=270,width=580pt}}
\end{center}
\caption{Delayed ADC versus prompt ADC pulse height 
in the liquid scintillator
showing the separation between neutrons (lower band) and photons
(upper band).}
\label{fig6}
\end{figure}

\begin{figure}
\begin{center}
\mbox{\epsfig{file=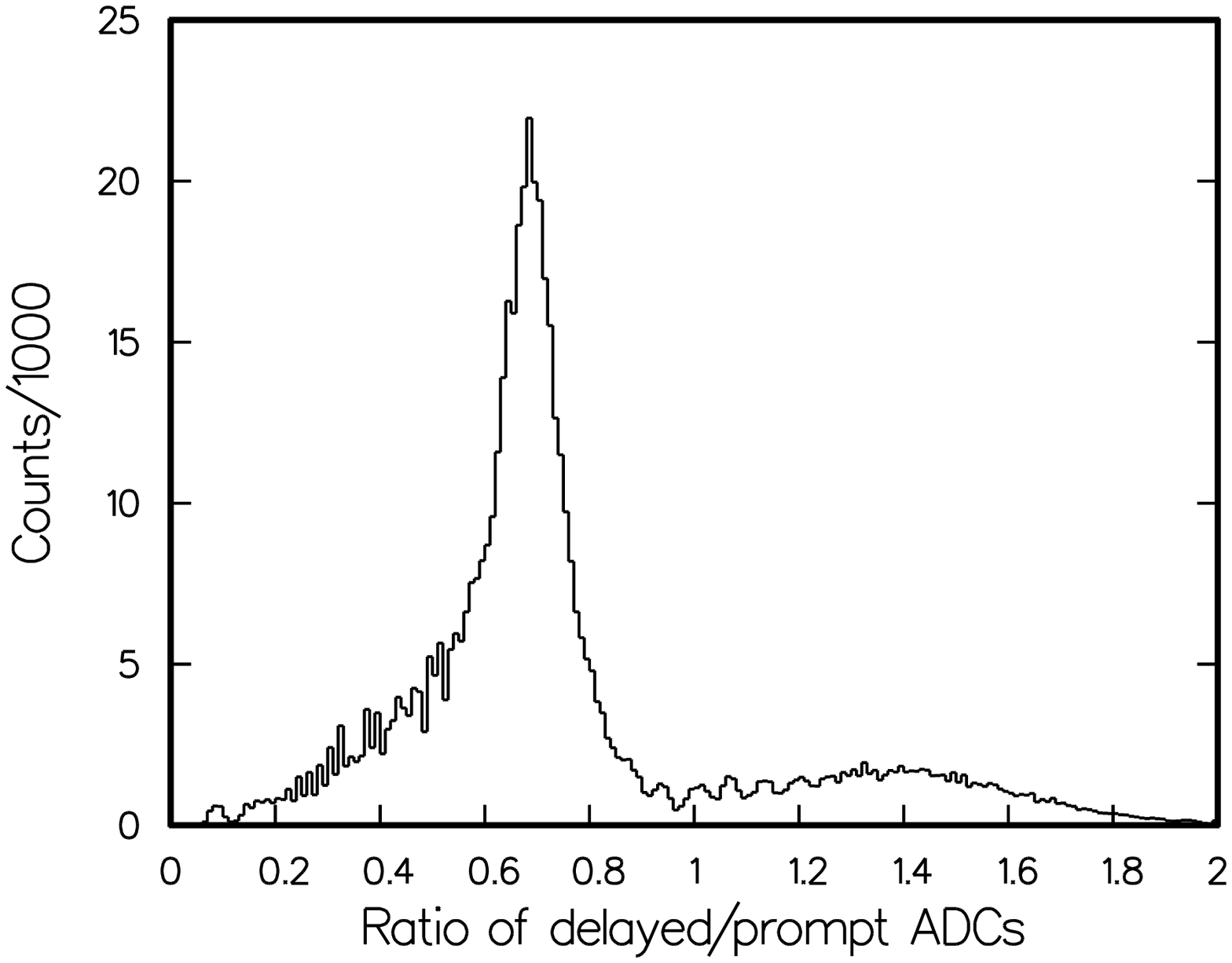,angle=270,width=580pt}}
\end{center}
\caption{Ratio of delayed/prompt ADC 
pulse height measurements used to
differentiate photons and neutrons in the liquid scintillators.}
\label{fig7}
\end{figure}

Once  neutron events were
identified in the liquid scintillator, 
the pulse height spectrum was unfolded to give the neutron
energy spectrum
using the program {\small FERD}~\cite{refferd}. 
That spectrum, from 1~to 10~MeV, is
compared with the spectrum measured by the Bonner spheres
over the same energy range in Fig.~8 for beam stop configuration~3.
Table~\ref{table:lstable} gives a comparison of these spectra
for four beam stop configurations.  They
show variations
similar to those seen with the $^3$He counters in Table~\ref{table:He3table},
again suggesting consistency at the factor of two level.

\begin{figure}
\begin{center}
\mbox{\epsfig{file=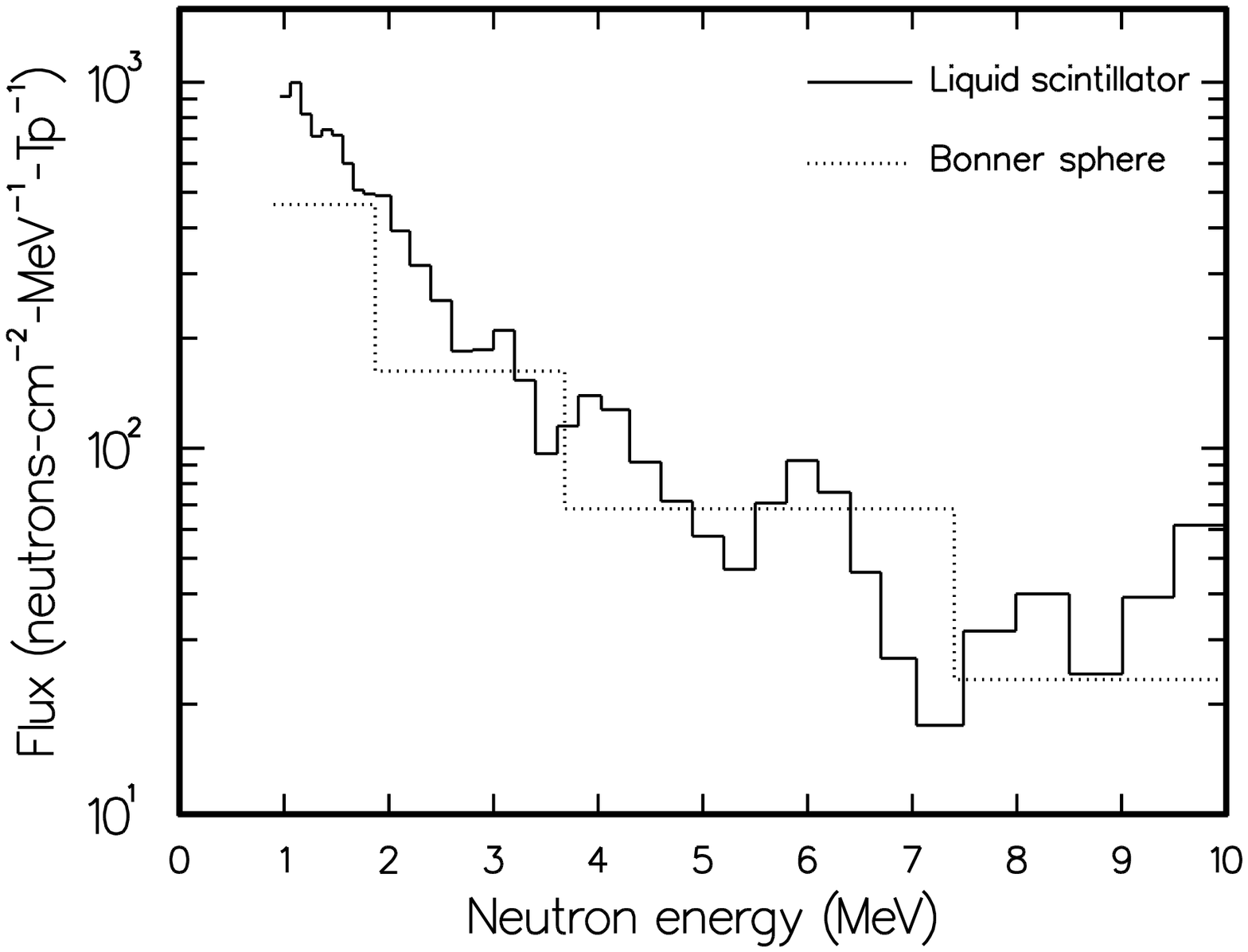,angle=270,width=580pt}}
\end{center}
\caption{Comparison of neutron energy spectra from
Bonner spheres and liquid scintillators.}
\label{fig8}
\end{figure}

\begin{table}
\begin{center}
\begin{tabular}{ l  c  c  c  c }
\hline
Configuration & 1 & 2 & 3 & 4 \\
         & No beam & Tungsten & Shielded  &   Extra   \\
         &   stop  &   only   & beam stop & poly, Pb  \\
\hline
Upstream position  & 0.4 & 0.5 & 0.6 & 1.9 \\ 
Downstream position  & 1.7 & 0.9 & 0.9 & 1.0 \\ \hline
\end{tabular}
\end{center}
\protect \caption{
The ratio of the neutron flux from the liquid scintillator to that from
Bonner spheres
 integrated over the range 0.9 to 7.4~MeV. 
}
\label{table:lstable}
\protect
\end{table}

  The liquid scintillator was also used to measure the photon flux and
energy spectrum. Events with  pulse shapes consistent with minimum
ionizing particles were assumed to be
originated by photons if there was 
no signal within $\pm$10~ns 
in the 10.2 cm $\times$ 10.2 cm $\times$ 0.32 cm plastic
scintillators located 9~cm upstream and downstream from the center of
the liquid scintillator.
The number of counts due to photons was approximately five times
greater than that due to neutrons, as shown in Fig.~9. 
The neutron spectrum showed no
structure, but there were two Compton edges visible in the photon
spectrum, one near 35 counts and the other near 250 counts.
These are the result of  photons with energies near 0.5~MeV and 
2~MeV, respectively,
and correspond to a combination of photons from the capture of 
neutrons on boron (0.477~MeV) and
positron-electron annihilation (0.511~MeV), which are too close in energy to
be resolved by this detector, and
the $^1$H($n,\gamma$)$^2$H reaction (2.2~MeV).
The photons from these sources were a fairly small fraction of the
total photon rate. 
The liquid scintillator was not sensitive to photons with energies below
about 100~keV.  
The discriminator threshold 
was set near that
level because the pulse shape discrimination method does not work at
lower energies.  
 
\begin{figure}
\begin{center}
\mbox{\epsfig{file=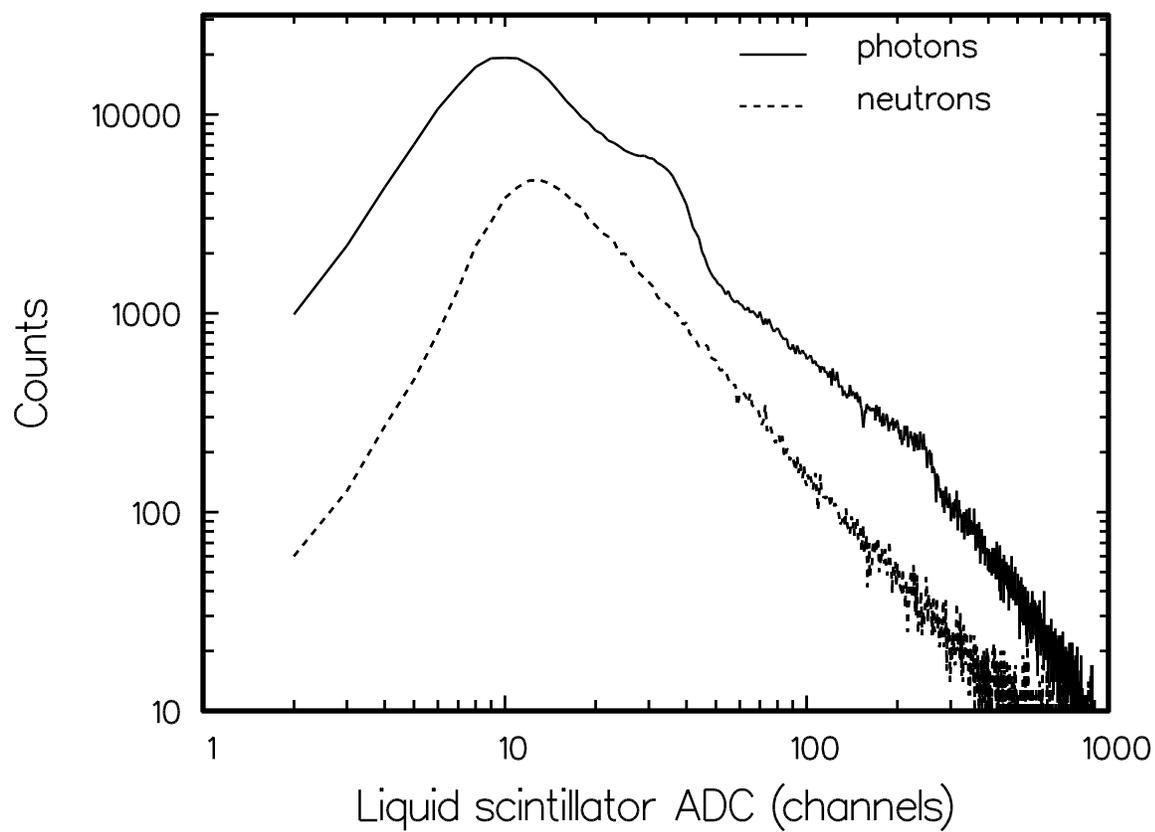,angle=270,width=580pt}}
\end{center}
\caption{ADC spectra from liquid scintillator for 
events identified as photons or neutrons.  The suppression of counts
below 10~ADC channels is due to an instrumental threshold.}
\label{fig9}
\end{figure}

An intrinsic germanium
detector was used to identify photons that were produced from neutron
capture.   
The germanium detector results are consistent with those from the liquid
scintillator. 
For illustration, two 
spectra taken with the germanium detector
are shown in Figs.~10(a) and 10(b).
For the spectrum in Fig.~10(a), the gain of the 
amplifier
was set low
to collect data on high energy photons. 
For the spectrum shown in Fig.~10(b), the gain was raised
to be sensitive to low energy photons. Several photon lines are visible,
but  they did not constitute a large fraction of the overall rate.
Some of the visible lines 
are: the 7.4~MeV line
from neutron capture on Pb; the 2.2~MeV line from radiative
capture on hydrogen; the positron-electron annihilation line at 511~keV;
the Doppler broadened
peak at 477~keV from the
decay of an excited $^7$Li that is produced along with an $\alpha$ by neutron 
capture on
$^{10}$B; and several lines (59, 78, 146, and 201~keV)
from capture on $^{186}$W.
The tungsten lines in Fig.~10(b) are visible because for this data the
beam stop 
did not have  its full complement of outer lead shielding.
With full lead shielding the tungsten lines disappear.

\begin{figure}
\begin{center}
\mbox{\epsfig{file=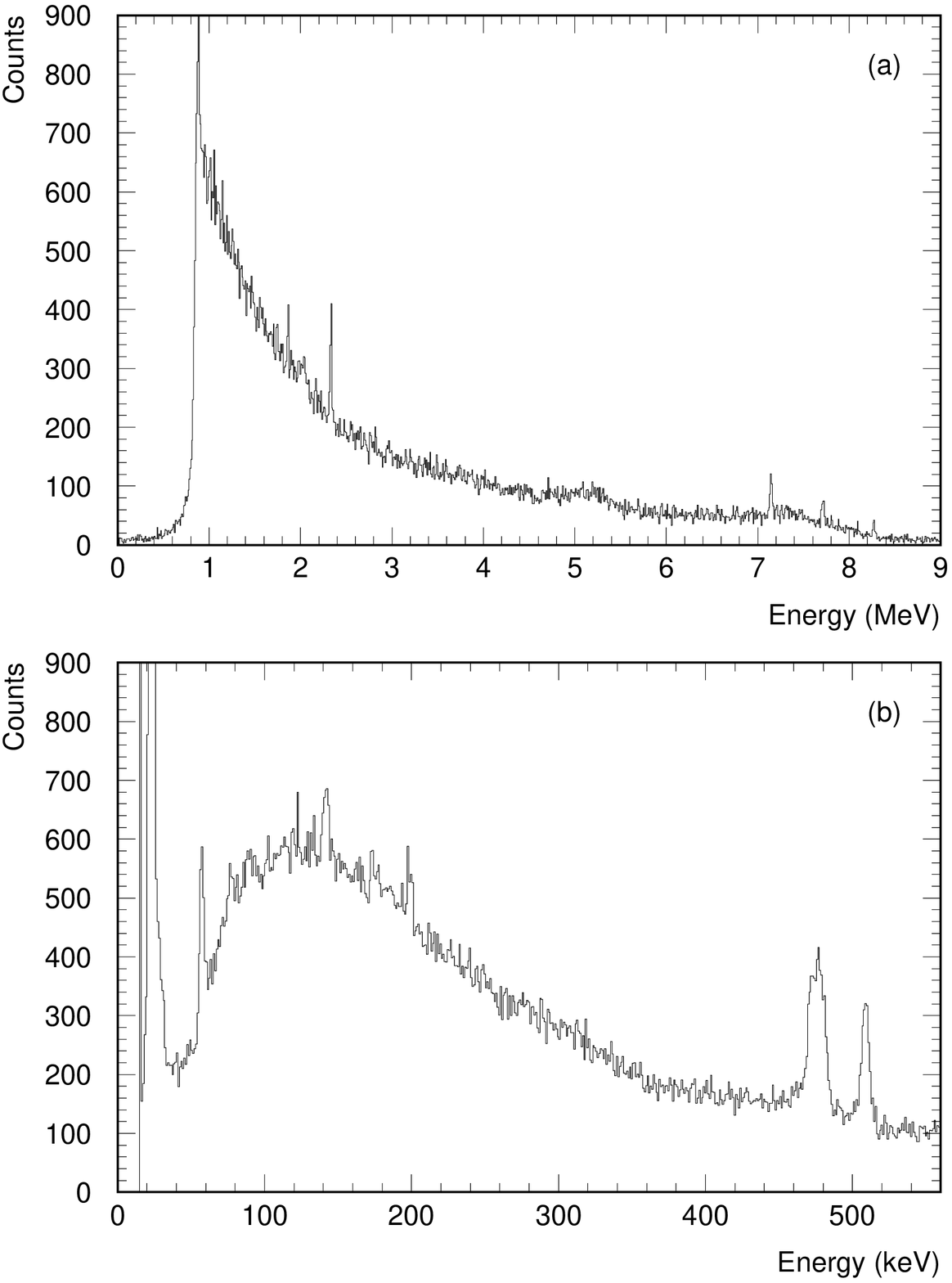,height=500pt}}
\end{center}
\caption{Sample energy spectra from the intrinsic germanium 
detector (a) with low amplifier gain and (b)
with high amplifier gain.}
\label{fig10}
\end{figure}

 Charged particles were identified using a coincidence between the
liquid scintillator and the plastic scintillators. The pulse shape from
the liquid scintillator was required to be consistent with a minimum
ionizing particle. Signals  from the plastic scintillators
were required to satisfy one of two conditions: either   both plastic
scintillators had signals with measured times within $\pm 10$~ns of
the liquid scintillator pulse, or  one of the signals had a
measured time
within $\pm 10$~ns of the liquid scintillator and also had  pulse 
height consistent with a minimum-ionizing
charged particle. 

  The relative fluxes of neutrons, photons, and charged particles
determined from the liquid scintillator and of neutrons
determined from the Bonner spheres are collected 
in Table~\ref{table:lsrates} for four beam stop configurations. 
The shielding added to the metallic beam stop
reduced the flux of neutrons and
photons, as measured by the liquid scintillators, by approximately a
factor of three when compared to the unshielded beam stop. The effect of
the shielding on the neutron flux as measured by the Bonner spheres was
much greater since that detector was sensitive to lower energy neutrons.
The shielding was not very effective in stopping the charged particle
leakage.

\begin{table}
\begin{center}
\begin{tabular}{ c c c  c  c  c }
\hline
 &  &  1 & 2 & 3 & 4 \\
   &     & No beam & Tungsten & Shielded  &  Extra   \\
Particle & Detector   &   stop  &   only   & beam stop & poly, Pb  \\ \hline
 &  & \multicolumn{4}{ c }{upstream position}  \\ 
neutrons & Bonner spheres & 1.0 & 430 & 80 & 19 \\ 
neutrons & liquid scintillator  & 1.0 & 85 & 29 & 16 \\ 
photons & liquid scintillator  & 1.0 & 19 & 7.6 & 4.6 \\ 
charged particles & liquid scintillator & 1.0 & 2.1 & 1.5 & 1.4 \\ 
 & & \multicolumn{4}{ c }{downstream position}  \\ 
neutrons & Bonner spheres  & 1.0 & 623 & 131 & 46 \\ 
neutrons & liquid scintillator & 1.0 & 63 & 28 & 19 \\ 
photons & liquid scintillator & 1.0 & 17 & 7.5 & 5.3 \\ 
charged particles & liquid scintillator & 1.0 & 2.4 & 2.4 & 2.5 \\ \hline
\end{tabular}
\end{center}
\protect \caption{
Relative particle flux  measured by the liquid scintillator and Bonner
spheres for  four beam stop configurations. 
The energy range covered for each particle type is decribed in the
text.} 
\label{table:lsrates}
\protect
\end{table}

\subsection{\normalsize{\it Drift Chamber Shielding Tests}}

Drift chambers, because of their close proximity to the beam stop,
were of special concern during the beam tests.
To investigate the types of particles causing drift chamber
hits, we operated a small test drift chamber that could be locally
shielded with polyethylene and lead.
The test drift chamber was an early prototype
of the E791 drift chambers and had
the same wire size and cell geometry, but had only a few wires
that were
17~cm long.
It had an aluminum frame with
Mylar windows on two sides. 
On four sides of the test
chamber (outside any shielding materials),
we mounted plastic scintillators to tag charged particles.
Coincidences between pairs of wires or between wire and scintillators  were
formed using commercial  logic units and were
counted using {\small CAMAC} scalers read out by the E791
data acquisition system. 

While the test chamber measurements do not lend themselves to precise
quantitative conclusions,
they provide general
indications of the major sources of drift chamber hits.  
Sample results
are given in Table~\ref{table:dctest} from data taken in
two different locations of the test drift chamber with the 
shielded beam stop
(configuration~3).  One location was immediately upstream of the beam stop and
occupied part of the space normally occupied by  the right-side
DC3 chamber.  
The other location was immediately downstream of the beam stop and
occupied part of the space normally occupied by  the left-side
DC4 chamber.  In other words, the test chamber was placed at the
locations of the E791 drift chambers which had been removed for the
purpose of allowing access to the beam stop.
The information displayed in Table~\ref{table:dctest} is in the form of
hit rates in various local shielding scenarios normalized to the hit rate
in the same chamber in the same location
with no local shielding.  
Table~\ref{table:dctest} also shows the fraction of  wire hits
accompanied by signals in the surrounding scintillators.
It should be recalled that
the charged decay products of
kaon decays in the beam constituted a significant source of hits.
This source would not be 
significantly affected by locally shielding the test chamber.  

\begin{table}
\begin{center}
\begin{tabular}{ l  c  c  c  c }
\hline
 & \multicolumn{2}{c }{Upstream position} & 
\multicolumn{2}{c }{Downstream position} \\ \hline
Local Shielding & Chamber & Fraction with & Chamber & Fraction with \\
description & rate  & scintillator &  rate   & scintillator  \\ \hline
No local shielding      & 1.00    & 0.38  & 1.00    & 0.69  \\ 
0.32 cm lead      & 0.81    & 0.38  & 0.89    & 0.71  \\ 
1.27 cm lead      & 0.64    & 0.37  & 0.87    & 0.81  \\ 
5 cm lead         & 0.47    & 0.37  & 0.59    & 0.80  \\ 
5 cm borated (5\%) polyethylene & 0.83    &  --   & 0.92    &  --   \\ \hline
\end{tabular}
\end{center}
\protect \caption{
Relative rates in a small test drift chamber, and the 
fraction of hits that deposited
energy in  plastic scintillators surrounding the chamber, for two
different locations as described in the text, with the beam stop in
configuration~3. 
All rates
are quoted with respect to the rate measured 
with no local shielding of the test chamber.
} 
\protect
\label{table:dctest}
\end{table}

One important conclusion can be drawn from the fact that
the test chamber rates decreased by only 17\%
in the upstream position, and  decreased by even less (only 8\%) in the
downstream location, when 5~cm of borated  polyethylene 
was placed around the test chamber.
This suggests that neutron interactions in the 
chamber gas comprise
a small fraction of the drift chamber rates in
configuration~3.  In other words, our shielding of the beam stop
succeeded in removing low energy neutrons as a major source of hits in the
drift chambers nearby.

Another conclusion from these measurements is that in the location immediately
downstream of the beam stop
chamber rates are caused mainly by fairly energetic charged particles.
This is indicated by the large fraction of chamber hits accompanied by
scintillator hits and the fact that 5~cm of lead shielding (between the
chamber and the scintillator) reduced this hit rate
by only 41\% while at the same time the fraction of hits accompanied by
scintillator hits increased.  Unfortunately, whether a large fraction of 
these penetrating charged 
particles originated in the beam stop, as opposed to in decays of beamline
$K^0_L$'s, cannot be inferred from these measurements.
In contrast, in the location upstream of the beam stop, chamber rates
decreased more steeply as lead shielding was added (reaching a 
53\% drop at 5~cm) while
the fraction of chamber hits accompanied by scintillator
hits did not increase.  Evidently, some mixture of soft
charged particles and photons  account for a large fraction
of the chamber hits in the location immediately upstream of the beam stop.

\subsection{\normalsize\it Cherenkov Phototube Shielding Tests}

A series of tests were performed to investigate the
Cherenkov counter rates in the presence of the beam stop.
We accumulated data using
four Cherenkov
photomultiplier tubes (PMT's),
which had been removed from the E791 Cherenkov
counter so  they could be shielded locally 
with polyethylene and lead.
For these tests, the beam-left Cherenkov
counter  was removed from the area, making room for the special
PMT test setup.  The location of the test setup is indicated in
Fig.~3 by the circled letter $C$.
The location of the test PMT's
was approximately the same as when 
attached to the Cherenkov counter gas
volume in E791.
On the other side of the beam, the beam-right Cherenkov
counter was operated normally.
The tests reported here were done with the beam stop in
configurations close to configuration~3.  Small changes in
beam stop shielding did not affect Cherenkov PMT rates, so that
some small variations in shielding were made in parallel with these
tests.

The Cherenkov PMT's were RCA/Burle model 8854 photomultiplier tubes, 
which have high gain and are
capable of detecting single photoelectrons.
Discriminator thresholds
were set low, so that they were
sensitive to single photoelectron pulses for the tests.
One PMT was connected to a clear Lucite
hollow cylindrical support that served in E791
to connect the PMT to the window of the
Cherenkov box. The PMT and Lucite support were
surrounded by a 0.64 cm thick iron cylinder to reduce the fringe field
from the downstream spectrometer magnet. 
The second PMT was identical to the first except that the
clear Lucite support was replaced by 
a black nylon support to test whether light
was generated in the Lucite. The third PMT had an iron
cylinder surrounding it that was twice as thick as normal. The fourth
PMT was set up to permit shielding on all sides with lead, polyethylene,
and {\small FLEX/BORON}. 
The rates in each of these PMT's
were measured using the standard E791 data acquisition system and the
same pseudo-random trigger described previously. 
 
The test PMT's had hit rates
only 30\% lower than the PMT's which viewed the gas volume.
It is likely that the hits
in the isolated PMT's were due mainly to low energy
photons and thermal neutrons.
Electrons from photons interacting in the glass
(or from nearby material if the electrons subsequently pass through
the glass)
could generate the
pulse height required to trigger the Cherenkov
discriminators.
Thermal neutrons could be captured
by the
$^{10}$B in the borosilicate glass from which the photomultiplier tube
was constructed. 
Presumably the rates on the PMT's on the Cherenkov counter were
derived from the same sources, plus the light generated by
kaon decay products above  Cherenkov threshold.
Results of the tests of the fourth PMT with
different types of local shielding are given in  Table~\ref{table:cer}.
Shielding the photomultiplier tube  with a
thin layer of {\small FLEX/BORON}, which stopped only slow neutrons,
reduced the rate by 24\%. Surrounding the PMT with 5~cm of lead and 5~cm of
borated (5\%) polyethylene eliminated two-thirds of 
the rate.
These results indicate that most of the Cherenkov hits induced by the
beam stop can be attributed to low energy photons and neutrons, which
can be substantially suppressed by locally shielding the PMT's.

\begin{table}
\begin{center}
\begin{tabular}{ l  c }
\hline
Shielding description & Normalized rate \\ \hline
No shielding                                 & 1.00    \\ 
0.64 cm {\small FLEX/BORON}                  & 0.76    \\ 
0.64 cm lead                                 & 0.79    \\ 
5 cm lead                                    & 0.58    \\ 
5 cm lead surrounded by & \\
\qquad 5 cm borated (5\%) polyethylene                    & 0.33    \\ \hline
\end{tabular}
\end{center}
\protect \caption{
Relative Cherenkov 
photomultiplier tube rates with several types of 
local shielding. Rates are
normalized to the rate for the PMT when unshielded.}
\protect
\label{table:cer}
\end{table}

\section{Monte Carlo Simulations}

We performed
Monte Carlo simulations to assist in the design of
the beam stop and to supplement the beam test 
measurements.
The initial series of Monte Carlo simulations, using the {\small HETC}
hadronic cascade Monte Carlo and  {\small MORSE}
low-energy neutron transport code (part of the
{\small CALOR89}~\cite{calor} package),
were performed
prior to the beam tests to provide input to the initial beam stop
design.
Subsequently, a version of the {\small CALOR89}
program~\cite{gcalor} coupled to {\small
GEANT}~\cite{geant} became available and was used
for a much more extensive series of simulations~\cite{SWthesis}.
Consistent results were obtained in both cases.
The more recent simulations are described in this paper.

  The {\small GEANT-CALOR} Monte Carlo package consists of several different
transport programs:

\begin{itemize}

\item The Nucleon-Meson Transport Code ({\small NMTC})~\cite{nmtc}
transports nucleons, charged pions, and muons. 
The
program transports protons in the energy range  1--3500~MeV;
the range for neutrons is 20--3500~MeV.
For   charged pions and muons
the energy range covered is  0.1-- 2500~MeV.

\item Monte Carlo Ionization Chamber Analysis Package 
(\small MICAP)~\cite{micap} transports neutrons below 20~MeV.

\item {\small GEANT}, which includes {\small
FLUKA}~\cite{fluka}, transports everything, including photons,
 that is not transported by
{\small NMTC} or {\small MICAP}. For nucleons and pions between the
{\small NMTC} cutoff and 10 GeV, a combination of {\small FLUKA} and
{\small NMTC} is used. 

\end{itemize}

The neutral beam used for these tests consisted primarily of neutrons with an 
average energy of about 8~GeV.  
Neutral kaons comprise about 4--8\% of the beam
and there
is also some contamination from photons.
In the simulations
discussed here, only the neutron component of the beam was treated.
The neutron momentum spectrum which was used as input for
the {\small GCALOR} simulation of interactions in the beam stop 
was generated using {\small GHEISHA}~\cite{gheisha},
normalized at a total flux of $3.7 \times 10^9$~n/Tp.
Charged particles from the decays of  $K_L^0$'s 
in the beam account for a large
fraction of rates in the detectors, of course, but that contribution does not
depend on the beam stop configuration.  Consequently,
when comparisons of Monte Carlo 
predictions were made to data, the comparisons were made to the difference of
measured rates between running configurations with a beam stop and
no beam stop.  In other words, contributions to rates from $K^0_L$ decays
were subtracted out using data taken without the beam stop in place.
Photons in the beam could be neglected because they do not contribute
significantly to leakage from the beam stop.

  The beam stop materials, spectrometer magnets, and shielding-block walls
of the experimental cave
were included in detail using the {\small GEANT} geometry package.
In all, 65 unique volumes consisting of up to 20 different materials were 
included in the simulation geometry.  
Neutron propagation was followed down to thermal energy. 
Neutrinos were ignored, but all other particles were tracked down
to 10~keV.
However, it was necessary to make some assumptions to simplify the
simulations.  Since the probabilities of neutrons and photons interacting
in drift chambers is very small, it would have been very inefficient
to simulate these interactions.
Instead we estimated the interaction probability for neutrons and
photons using
the particle energy and the amount and type of material traversed.
Then neutron and photon interaction rates in the drift chambers were
calculated based on the Monte Carlo generated neutron and photon
fluxes. 
The neutrons were assumed to interact with the   
hydrogen, carbon, and argon 
atoms  in the drift chamber gas  (a 50-50
mixture of argon-ethane).
The ($n,p$) cross section was taken to be 20~barns 
for neutrons with energies below 75~keV and 
(5.5~barns)/$\sqrt{E{\rm (MeV)}}$ 
for neutrons with energies above 75~keV~\cite{BNL325}. 
We did not include the
interactions of neutrons with energies below 10~eV since they could not
cause ionization and the capture cross section is small. 
The interaction of photons in the drift
chambers was approximated by parameterizing the interaction probability
as the thickness (22~mg/cm$^2$ per view)
divided by the attenuation length. The attenuation
length for low-Z materials (except hydrogen) was approximated as 
14.4~g/cm$^2\times \sqrt{E{\rm (MeV)}}$ in the energy range  
0.01--5~MeV~\cite{PDGgamma}.

\section{Comparisons of Data and Monte Carlo}

Comparisons between Monte Carlo predictions and
our measurements  are possible.  For the specialized detectors
which  measured neutrons and photons, it is possible to make a
direct comparision.  For E791 detectors, such as the drift chambers,
it is possible to compare only total rates, which requires summing 
Monte Carlo generated predictions from neutrons, photons, and charged
particles.  A third type of comparision is something of a hybrid of
these two.  It is possible to use the specialized detector measurements
of neutron, photon, and charged particle fluxes, supplemented by
Monte Carlo, to predict drift chamber rates.  This  comparision
depends on the Monte Carlo because the specialized detectors were not
in the same locations as the drift chambers and because the different
energy response of the drift chambers must be taken into account.
Another complication which applies to  comparisions of predictions
with measured drift chamber rates is that a single particle may have
caused several adjacent hits.  For example, a charged particle
may be incident at a large angle so that it traverses
more than one cell or a neutron may create (through the
interaction with a proton in the gas) extremely large ionization, which
might result in multiple hits via cross talk in the electronics.
We have attempted to account for these complications in the comparisions
which follow, but in these cases
some extra uncertainty
appertains.

The data obtained with the Bonner spheres
provided neutron energy spectra that are directly compared to the spectra from
the {\small GCALOR}  simulations  in Figs.~11 and 12.
Fig.~11 shows spectra obtained both upstream and downstream of the beam
stop for the metallic beam stop of configuration~2.
Fig.~12 shows the spectra obtained both upstream and downstream of the
beam stop for the fully shielded configuration~3.
In both Figs.~11 and 12, two sets of Monte Carlo generated neutron
spectra are shown.  The solid dots show the predicted neutron spectra
for simulations that include the full experimental geometry (i.e., includes
the effects of scattering on walls), while the open circles show only
neutrons emerging directly from the beam stop without further scattering.
Several features of these comparisons are evident:

\begin{figure}
\begin{center}
\mbox{\epsfig{file=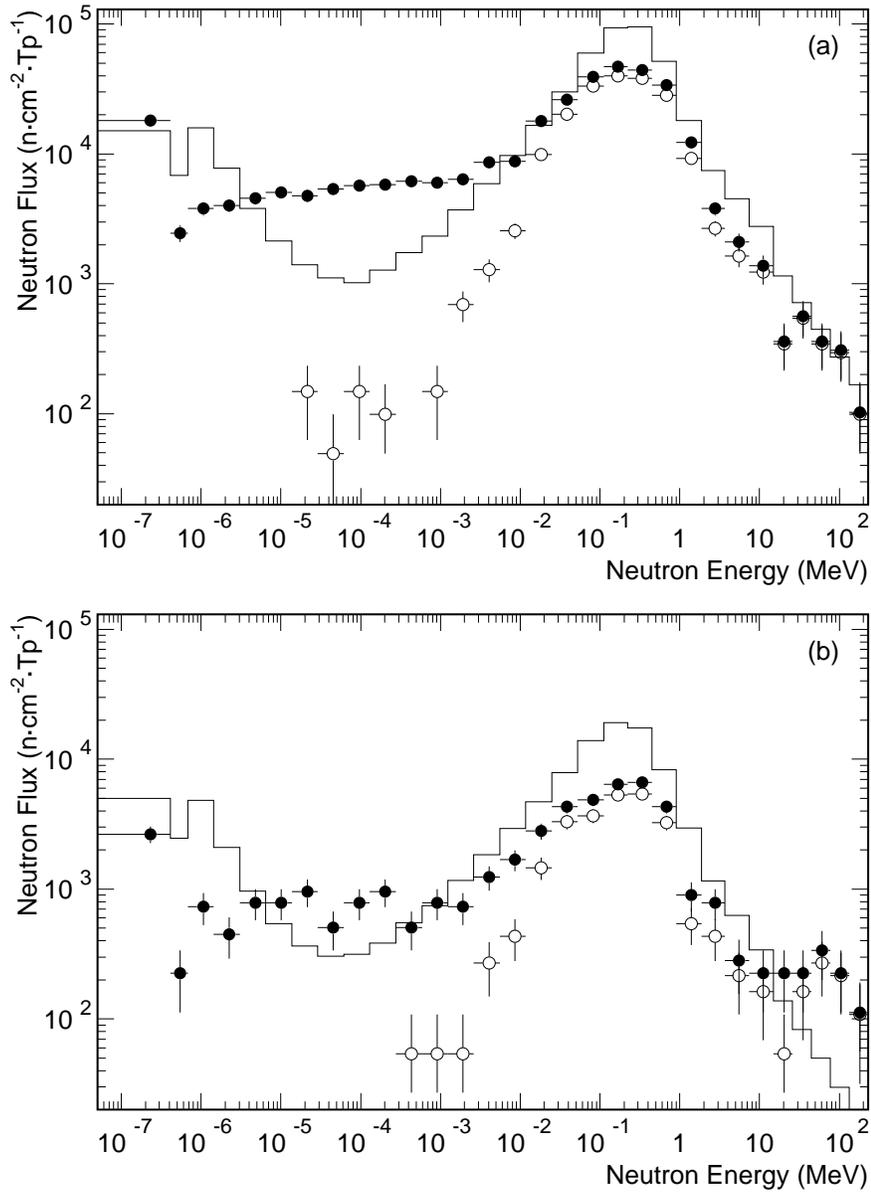,height=500pt}}
\end{center}
\caption{Neutron flux from
the Monte Carlo and the Bonner sphere measurements for the unshielded
metallic beam stop (configuration~2), 
both upstream (a) and downstream (b) of the
beam stop.  The solid line is the Bonner sphere result.  
The Monte Carlo predictions for neutrons emerging from the beam stop
are shown for two cases:
with (solid dots) and without (open cirles) rescattering.}
\label{fig11}
\end{figure}

\begin{figure}
\begin{center}
\mbox{\epsfig{file=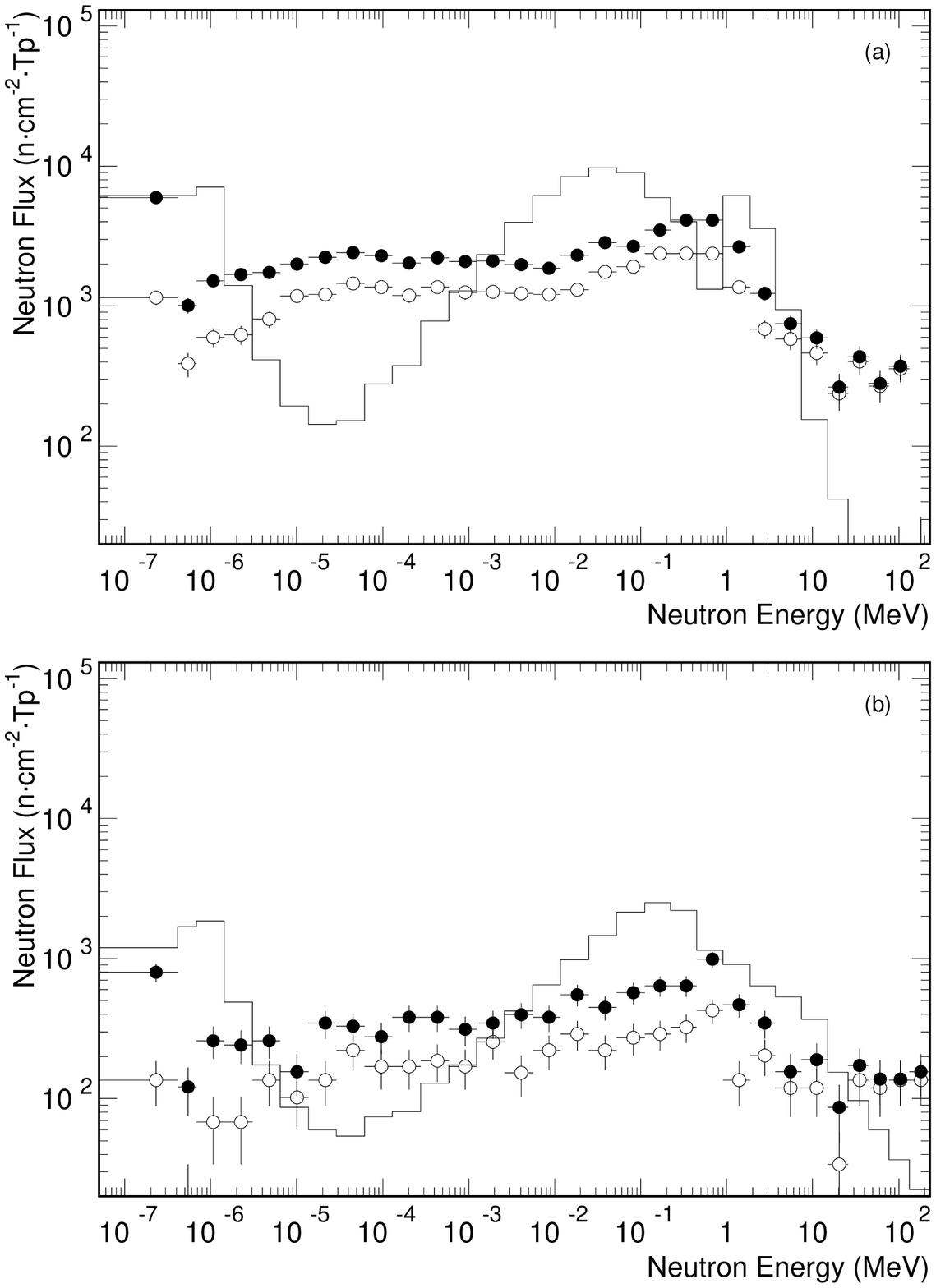,height=500pt}}
\end{center}
\caption{Neutron flux from
the Monte Carlo and the Bonner sphere measurements for the shielded
metallic beam stop (configuration~3), 
both upstream (a) and downstream (b) of the
beam stop.  The solid line is the Bonner sphere result.  
The Monte Carlo predictions for neutrons emerging from the beam stop
are shown for two cases:
with (solid dots) and without (open cirles) rescattering.}
\label{fig12}
\end{figure}

\begin{enumerate}

\item For the bare metallic beam stop (configuration~2), 
both Monte Carlo predictions
and the Bonner sphere neutron spectra peak in the energy range between
0.1 and 1~MeV.  The Bonner sphere fluxes are higher  by a factor of two to
three.  In the range between  1~MeV and 7~MeV where Bonner sphere
results overlap the liquid scintillator measurement, the liquid scintillator
indicated the lower neutron flux.

\item For the bare metallic beam stop (configuration~2),
most neutrons with energies below 0.01~MeV are the
result of rescattering after leaving the beam stop,
as can be seen from comparing the solid dots to open circles in
Fig.~11.
While neutron leakage from the shielded configuration 
(configuration~3 in Fig.~12) is significant
at energies below 0.01~MeV,  the fraction of neutrons
below about 1~MeV which have rescattered is at least one-half
and increases significantly for lower energies.
Rescattered neutrons dominate at thermal energies.
These results emphasize the importance of including the external geometry
in the simulation for a practical situation.

\item In the energy range of 0.1 to 1~MeV where the neutron leakage from
the beam stop is the most severe, both the Monte Carlo and the
measured spectra indicate that the shielding added to the beam stop
for configuration~3 reduced the leakage by 
roughly an order of magnitude.

\item The dip in the Bonner sphere spectra between about 1~eV and 1~keV
is not reproduced in the Monte Carlo.  As noted earlier, this feature
is probably an artifact of the unfolding method used to obtain
the neutron energy spectrum from the Bonner sphere measurements.

\end{enumerate}

  A similar comparison with the Monte Carlo was made for the photon
spectrum measured by the liquid scintillator. The photon spectrum was
calculated from the measured pulse height assuming all photon
interactions were due to Compton scattering. The comparison,
shown in Fig.~13 for configuration~3,
indicates a similar energy 
spectrum, but the measurement gives
about a 30\%  
higher photon flux.

Ultimately, our goal is to
understand the cause of the hit rates in detectors located
near the beam stop, particularly the drift chambers since they are
the closest.  If the hit rates were excessive, this understanding
would help to further optimize the beam stop to reduce them.
To that end, we have compared drift chamber hit rates to the Monte
Carlo predictions, and to predictions based on  
the specialized detector measurements
supplemented by Monte Carlo.
As noted earlier, 
several complications must be taken into account in making such
comparisons:

\begin{figure}
\begin{center}
\mbox{\epsfig{file=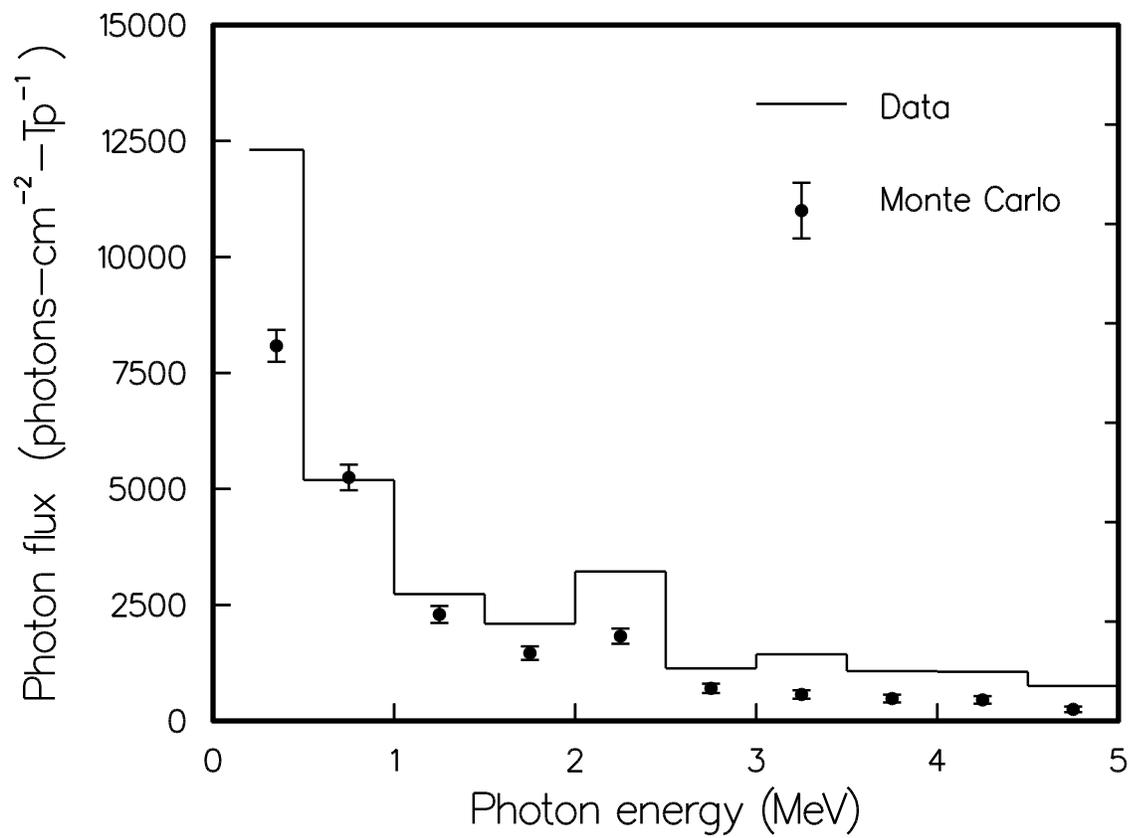,angle=270,width=580pt}}
\end{center}
\caption{Comparison of the photon flux predicted by
the Monte Carlo simulation 
and the liquid scintillator measurements upstream of the
shielded beam stop. }
\label{fig13}
\end{figure}

\begin{enumerate}

\item  Some of the drift chamber rate was due to cases in which a single
particle caused hits in many adjacent cells. To account for this,
the drift chamber rates were
recalculated in units of ``clusters'', defined as a group of hit
adjacent
wires  consistent with the interaction of a
single particle. 

\item The cluster rates due to particles from $K_L^0$ decays
upstream of the beam stop were subtracted using data taken with
no beam stop in place. 

\item The results from the liquid scintillator and Bonner sphere
measurements were adjusted to account for the angle at which
the neutrons and photons traversed the drift chambers. No angular
information was available from these detectors so a scale factor
was used based on the Monte Carlo calculations. 

\item The neutron and photon detectors did not occupy the same location as
the drift chambers so a correction was calculated from the Monte Carlo
to account for the position dependence of particle-induced rates.
In particular, for the downstream setup, the neutron and photon detectors
were between DC4 and DC5.  A correction was calculated to
effectively relocate the detectors to the position of DC4
 (i.e., to move
the detectors upstream by about 92~cm). 

\item A correction was applied to account for the energy thresholds in
the liquid scintillator. The liquid scintillator did not detect photons
with energies below 200 keV or charged particles that could not
penetrate the 1.2 g/cm$^2$ of material surrounding it. Both corrections
were calculated using the  Monte Carlo. 

\end{enumerate}

The correction factors for items 3, 4, and 5
were estimated separately for charged particles,
neutrons, and photons.  When combined,
the magnitudes of the corrections for the upstream
detectors were about 50\% for neutron-induced
rates, 70\% for photon-induced rates, and 120\% for charged particles.
In the downstream position, the magnitudes were roughly doubled because
of the position correction.
Table~\ref{table:dcall} shows the beam stop induced drift chamber
rates (for chambers DC3 and DC4)
for three beam stop configurations after the corrections are
applied. 
The normalization to the telescope at the target has been 
used
since it is the only normalization available for the liquid scintillators and
Bonner sphere measurements.  The rates are given in terms of clusters, as
defined above, per telescope hit.  
There were approximately $12.5 \times 10^3$
telescope hits/Tp on target.

\begin{table}
\begin{center}
\begin{tabular}{ l  c  c  c }
\hline
Configuration & 2 & 3 & 4 \\
         & Tungsten & shielded  &  extra   \\
         &   only   & beam stop & poly, Pb \\ \hline 
 & \multicolumn{3}{ c }{upstream position} \\ 
\underbar{Drift chamber rate (DC3)} & {\bf 3320} & {\bf 770} & {\bf 550} \\ 
\underbar{Monte Carlo} & & & \\
\ \ \ charged particles &
 2000 & 440 & 300 \\ 
\ \ \ neutrons &
 380 & 70 & 26 \\ 
\ \ \ photons &
 840 & 86 & 120 \\ 
\ \ \ {\bf Sum} &
 {\bf 3200} & {\bf 600} & {\bf 450} \\ 
\underbar{Estimates based on specialized detector measurements} & & & \\
\ \ \ charged particles &
 1340 & 380 & 240 \\ 
\ \ \ neutrons &
 370 & 80 & 16 \\ 
\ \ \ photons &
 270 & 110 & 63 \\ 
\ \ \ {\bf Sum} &
 {\bf 1980} & {\bf 570} & {\bf 320}  \\ \hline 
 & \multicolumn{3}{ c }{downstream position} \\ 
\underbar{Drift chamber rate (DC4)} & {\bf 2540} & {\bf 654} & {\bf 624} \\ 
\underbar{Monte Carlo} & & &\\
\ \ \ charged particles &
 860 & 380 & 240 \\ 
\ \ \ neutrons &
 210 & 39 & 16 \\ 
\ \ \ photons &
 170 & 73 & 87 \\ 
\ \ \ {\bf Sum} &
 {\bf 1240} & {\bf 490} & {\bf 340} \\ 
\underbar{Estimates based on specialized detector measurements} & & & \\
\ \ \ charged particles &
 1040 & 750 & 850 \\ 
\ \ \ neutrons &
 270 & 46 & 16 \\ 
\ \ \ photons &
 280 & 110 & 91 \\ 
\ \ \ {\bf Sum} &
 {\bf 1590} & {\bf 910} & {\bf 960}  \\ \hline
\end{tabular}
\end{center}
\protect \caption{ 
Comparison of measured E791 drift chamber rates (for DC3 and DC4)
from beam stop  (with magnet off) to those 
predicted by  Monte Carlo simulations and those predicted based on
measurements with liquid scintillator and
Bonner spheres.
Rates from $K_L^0$ decays were subtracted
from the data.  Rates are quoted in units of clusters per telescope hit,
as explained in the text.  
} 
\protect
\label{table:dcall}
\end{table}

  The results in Table~\ref{table:dcall} indicate:

\begin{itemize}

\item Absolute predictions of drift chamber rates
from both the Monte Carlo and estimates of 
particle fluxes from the specialized
detectors agree with the measured rates, and each other, 
typically to within a factor of two.

\item Both the Monte Carlo simulations and the specialized neutron
and photon detector measurements indicate that drift chamber rates
are dominated by charged particles, rather than neutrons or photons,
for all configurations of the beam stop.
This is an interesting result, and consistent with the indications
from the test drift chamber studies (section~4.4).
It should be recalled that the choice
of tungsten for the core material was based partly on the fact that
tungsten has a large inelastic neutron cross section down to 1~MeV.
Neutron leakage would be a larger problem with other core materials.

\end{itemize}

The fractional variation of actual drift chamber rates 
between different beam stop configurations
tends to be approximately reproduced in both the Monte Carlo predictions and
the estimates based on the neutron and photon detectors.
The exception was the downstream
measurement after additional shielding was added (configuration~4).
The Monte Carlo rates
dropped with the extra shielding while the drift chamber rate and the
rate calculated from the neutron and photon detector measurements stayed 
about the same. 
Also, the special
detector measurements indicated too high a rate
from charged particles  downstream of the shielded beam stops. 
These detectors, however, were intended for neutron and photon measurements.
The importance of charged particles became clear as a result
of these measurements, but were not a focus initially.

Other indications that charged particles were a major source of detector
rates were present in both the data and simulations.
In one of the many variations of the beam stop configuration 
that were tested, we added copper plates to the outside surface on
one side.
We found a roughly 25\%  drop in
rates in the nearby drift chamber with the addition of 5~cm of copper.
Also, left--right asymmetries appeared in the hit rates for
the drift chambers downstream of the beam stop
when
magnetic fields were present.
The {\small GCALOR} simulation exhibited qualitatively similar 
asymmetries which were due to protons 
escaping the beam stop.  These protons originated in collisions
with tungsten nuclei and usually escaped with 
momenta in the range of about
200~MeV up to 1~GeV.

\section{Discussion}

Subsequent to the tests described here, a number of changes were made
to the beam stop design.  For the most part they were intended to 
reduce the leakage of charged particles, since
our studies indicated that charged particles were the main source
of drift chamber rates near the beam stop.
In particular, 
more high-mass material was used in a number of locations.
Most of the borated (0.5\%)
polyethylene (density 0.84~g/cm$^3$)
on the sides of the beam stop was replaced with borated
(0.5\%) zirconium hydride polyethylene~\cite{rxe}
with a density of 3.6~g/cm$^3$.
This material has the attractive features of borated polyethylene,
good hydrogen density along with boron to capture thermal neutrons,
but is much more dense and has more stopping power for charged particles.
Also,
three of the polyethylene sections in the tunnel were replaced by copper
and the thickness of the outer lead layer was quadrupled.
The final beam stop as used in E871 is shown in Fig.~14.

\begin{figure}
\begin{center}
\mbox{\epsfig{file=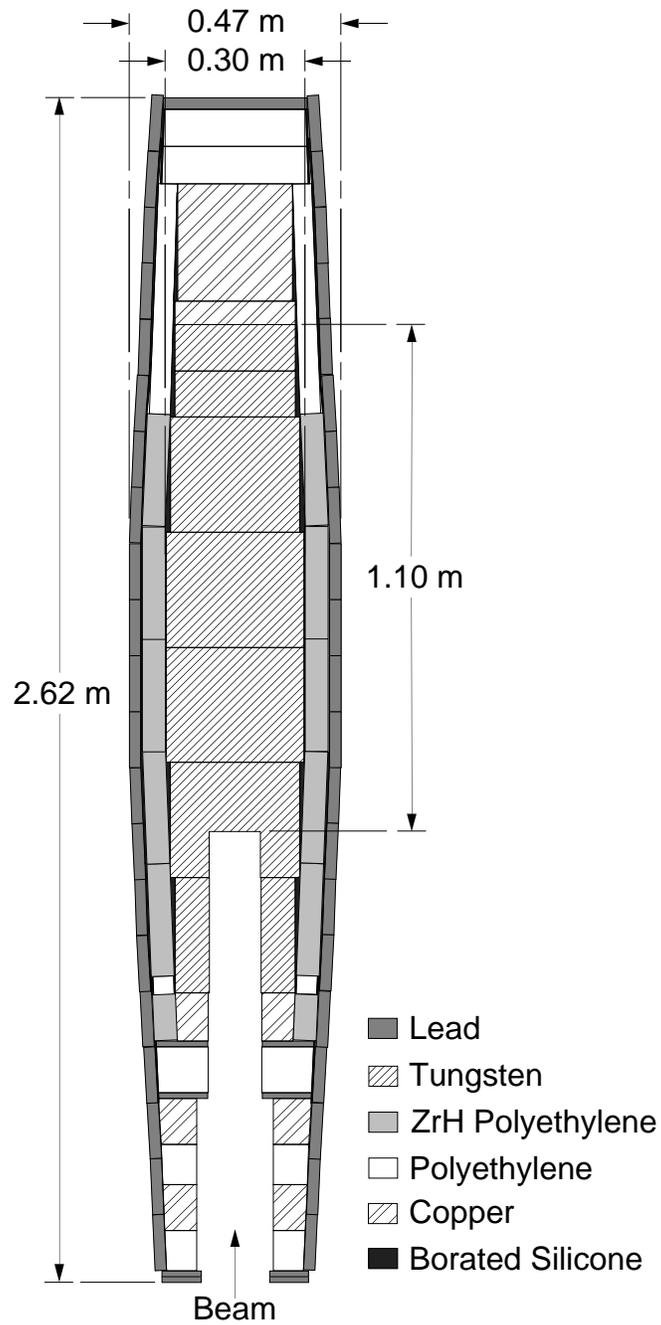,height=500pt}}
\end{center}
\caption{Horizontal cross section view
through the center
of the beam stop used in E871.
A thin layer of {\small FLEX/BORON} surrounded the borated zirconium hydride
polyethylene.}
\label{fig14}
\end{figure}

In addition to modifications to the beam stop, 
we decided for E871 to run with a platinum target at
a 3.75~degree production angle.  This change from the initial E791
configuration of the beam (a copper target with a 2.75~degree
production angle) 
was motivated by a number of
considerations.  The larger angle results in a more
favorable $K^0_L/n$ ratio.
The average neutron energy is also 
reduced, leading to less energy per neutron
deposited in the beam stop.  The platinum target is shorter
(for the same number of interaction lengths) than
the copper target, which benefits background rejection in the 
$K_L^0 \rightarrow \mu^\pm e^\mp$ search.
This improvement results from the fact that the target is oriented
at an angle with respect to the neutral beam and a shorter target presents
a smaller (i.e., more pointlike) kaon source as viewed by the
spectrometer.  For the purpose of studying target issues,
a small amount of test data was taken with a platinum target at
3.75~degrees near the end of the 1992 run in one beam stop
configuration, but that data was not useful for beam stop studies.

With the installation of E871 we replaced the drift chambers used during
the beam tests with straw chambers, making a direct
comparision between chamber rates with the final beam stop
design and the earlier design impossible.  
Having reached what we believed was a satisfactory design, we did
not undertake further beam tests.  E871 had successful physics runs
in 1995 and 1996.  While we did not conduct further beam tests,
we can estimate the contribution of the beam stop to straw chamber
rates by comparing the distribution of chamber hits measured with our
pseudo-random (pulser) trigger to that expected from $K^0_L$ decay
products from Monte Carlo.  
This estimate is straightforward because our measurements showed that
hits associated with the beam stop tend to be distributed almost
flat across the chambers.
In the worst-case
straw chambers --- those immediately upstream of the beam stop, 
our conclusion is that
$K^0_L$ decays account for about half of the rate averaged across
the chamber.
In the wires with the highest singles rate, about 600~kHz near 
inner (beam-side) edge
under typical data-taking conditions, 
$K^0_L$ decays account for about 70\% of the hits.

The plan to put a beam stop near active detectors in BNL E871 was
initially greeted with skepticism based on the conventional wisdom
that leakage, particularly of neutrons, would cause  untenably high
hit rates in the nearby detectors.
However, with the final beam stop in place,
E871 has run  at beam intensities
up to 20~Tp/spill (above our design goal of 15~Tp/spill)
without adverse rate effects in the chambers
nearest the beam stop.  Whether the successful 
beam stop development will have importance
for other experiments is not clear,  but it at least  indicates
that a hadron absorber can probably be integrated into many experimental
setups if appropriate care is taken in the design
and testing.

Our program of design and testing involved  significant beam tests
conducted over a two year period at the AGS and was supplemented
by extensive Monte Carlo simulations using  state-of-the-art
hadronic shower and neutron transport codes.
While ultimately successful, it is clear in retrospect that many
parts of this program could have been improved.
For example, the largest uncertainty in our measurements came from
our relatively poor monitoring of the neutral beam flux incident
on the beam stop.  It is difficult to monitor a neutral
beam, but in order to reliably compare different beam stop configurations
it is essential to know the incident flux and spectrum.
Our comparisons were never reliable to better than
20\%  owing to beam flux uncertainties, so that our optimization
of the beam stop may have been compromised by an inability to
recognize improvements between alternative configurations when they
were  smaller than 20\%.

Another limitation clear in retrospect was our marginal ability to detect
and identify
charged particles.  Our efforts in preparing 
specialized detectors were motivated by  concerns over neutron
leakage from the beam stop and over photons which we expected to
arise as  by-products of neutron capture.
Ultimately our conclusion is that  the increase in hit rates
in drift chambers caused by the 
fully shielded beam stop is dominated by charged particles,
which we were ill-equipped to measure.

While in practice we used Monte Carlo simulations in a limited way to
optimize the beam stop design, primarily by comparing the leakage
from slightly different configurations when it was not practical
to make measurements, one can ask what our measurements show regarding
the validity of the Monte Carlo programs used.
It is important to remember that our measurements were not planned
with the goal of validating or testing Monte Carlo programs, and if that
had been the goal, very different measurements would have been made.
Nonetheless, some general conclusions can be drawn with respect to
the simulation of a complex shielding geometry.

The {\small GCALOR}
Monte Carlo predictions seem to generally agree with our measurements
within a factor of two.
The differences are larger than the uncertainties in the incoming
beam fluxes can account for, although those uncertainties are significant
on this scale. 
Where there are disagreements in absolute numbers,
the shapes of spectra tend to agree reasonably well.
In view of the number and complexity of the
processes being simulated, which span the energy range from above 10~GeV
down to below 1~eV,
this level of agreement seems rather good.
While our measurements cannot distinguish between particles coming
directly from the beam stop and those which have rescattered on other
material, it is clear  that including the surrounding material 
in the Monte Carlo geometry was very important. 
The Monte Carlo programs are clearly powerful tools for evaluating the
efficacy of alternative absorber configurations, but should be 
used in concert with actual measurements whenever possible.

\vspace{1cm}

\noindent {\bf Acknowledgments}

We are grateful to T.A. Gabriel, B.L. Bishop, K. Furuno, T.
Jenkins, and W.R. Nelson for their help with running the {\small
CALOR} Monte Carlo, and C. Zeitnitz for his help with
{\small GCALOR}. We thank R. Brown, J. Scaduto,  
and the staff at the AGS
for their assistance during the beam stop construction. 
We appreciate the contributions of R. Erbacher, 
C. Hoff, N. Mar,
R. Martin, and M. Pommot-Maia in
collecting our data.  Most of the 
Monte Carlo simulations reported here were
run on the Physics Detector Simulation Facility (PDSF) 
at the SSC Laboratory.
This work was
supported in part by the U.S. Department of Energy, the National Science
Foundation, and the Robert A. Welch Foundation.

\clearpage


\newpage


\begin{thebibliography}{99}


\bibitem{proposal} A.P. Heinson {\it et al.}, BNL Proposal 871 (1990).

\bibitem{SGWmemo} S. G. Wojcicki, BNL E791 technical note KL-290 (1990);

\bibitem{SWthesis} S.D. Worm, Design and Evaluation of a Beam Stop in the
Spectrometer of a Rare Kaon Decay Experiment, Ph. D. thesis,
University of Texas at Austin (1995).

\bibitem{mumu} A.P. Heinson {\it et al.}, Phys.\ Rev.\  D 51 (1995) 985.

\bibitem{Vienna} S. Graessle {\it et al.},  
Nucl.\ Instr.\ and Meth.\   A 367 (1995) 138.

\bibitem{AnnsNIM} A.P. Heinson and D. Rowe, Nucl.\ Instr.\ and Meth.\ 
 A 321 (1992) 165. 

\bibitem{E605} Y. Hsiung {\it et al.}, Phys.\ Rev.\ Lett.  55 (1985) 457;
J. Crittenden {\it et al.}, Phys.\ Rev.\ D  34 (1986) 2584.

\bibitem{calor} T.A. Gabriel {\it et al.},
ORNL Report TM-11185 (1990).

\bibitem{rxe} {\small FLEX/BORON},
borated zirconium hydride polyethylene, 
and most other neutron shielding materials
were obtained from Reactor Experiments, Inc., 1275 Hammerwood Avenue,
Sunnyvale CA 94089.


\bibitem{neutr} E.C. Milner, BNL E791 technical note KL-256 (1989);
P. Buchholz and W.K. McFarlane, BNL E791 technical note KL-268 (1989).

\bibitem{MDpreprint} J. McDonough, Proceedings of DPF~92 (Fermilab, 1992),
World Scientific, p.\ 1708;
M. Diwan, SSCL-Preprint-192 (1993).

\bibitem{refbs} R.L. Bramblett, R.I. Ewing, and T.W. Bonner, Nucl.\ Instr.\
and Meth.\  9 (1960) 1.

\bibitem{refbunk} K.A. Lowry and T.L. Johnson, NRL Memorandum Report 5340
(1984).

\bibitem{refhe} K.-H. Beimer, G. Nyman, and O. Tengblad, Nucl.\ Instr.\ 
and Meth.\ A 245 (1986) 402.


\bibitem{refpsd} C.L. Morris {\it et al.},
Nucl.\ Instr.\ and Meth.\  137 (1976) 397.

\bibitem{refferd} B.W. Rust, D.T. Ingersoll, and W.R. Burrus, ORNL/TM-8720
(1983).


\bibitem{gcalor} C. Zeitnitz and T.A. Gabriel, Nucl.\ Instr.\ and Meth.\ 
A 349 (1994) 106. 


\bibitem{geant} R. Brun, F. Bruyant, M. Maire, A.C. McPherson, P. Zanarini,
CERN DD/EE/84-1 (1987).

\bibitem{nmtc} W.A. Coleman and T.W. Armstrong, ORNL-4606 (1970);
R.G. Alsmiller, Jr., F.S. Alsmiller and O.W. Hermann,
Nucl.\ Instr.\  and Meth.\  A 295 (1990) 337.

\bibitem{micap} J.O. Johnson and T.A. Gabriel, ORNL/TM-10196 (1987). 

\bibitem{fluka} P.A. Aarnio {\it et al.},
CERN Report TIS-RP/190 (1987).

\bibitem{gheisha} H. Fesefeldt, Aachen Technical Report PITHA 85-02, 
(September,1985).

\bibitem{BNL325} Cross Section Evaluation Working Group, ENDF/B-VI Summary
Documentation, Report BNL-NCS-17541 (ENDF-201) (1991), edited by 
P.F. Rose, National Nuclear Data Center, Brookhaven National Laboratory,
Upton, NY, USA.

\bibitem{PDGgamma} R.M. Barnett {\it et al.},
Review of Particle Physics, Phys.\ Rev.\ D 54 (1996) 140.

\end{thebibliography}
\end{document}